\documentclass[twocolumn]{aastex631}

\usepackage{graphicx}

\received{2024 September 4}
\revised{2025 February 18}
\accepted{2025 March 27}

\shorttitle{Evolution of Neutral Hydrogen}
\shortauthors{CHILES Collaboration}

\begin{document}

\title{CHILES VIII: Probing Evolution of Average HI Content in Star Forming Galaxies over the Past 5 Billion Years}

\correspondingauthor{Nicholas Luber}
\email{nicholas.m.luber@gmail.com}

\author{Nicholas Luber}
\affiliation{Department of Astronomy, Columbia University, Mail Code 5247, 538 West 120th Street, New York, NY 10027}

\author{D.J. Pisano}
\affiliation{Department of Astronomy, University of Cape Town, Private Bag X3,
Rondebosch 7701, South Africa}

\author{J.H. van Gorkom}
\affiliation{Department of Astronomy, Columbia University, Mail Code 5247, 538 West 120th Street, New York, NY 10027}

\author{Julia Blue Bird}
\affiliation{National Radio Astronomy Observatory, P.O. Box O, Socorro, NM 87801, USA}

\author{Richard Dodson}
\affiliation{International Centre for Radio Astronomy Research (ICRAR), University of Western Australia, 35 Stirling Hwy, Crawley, WA 6009, Australia}

\author{Hansung B. Gim}
\affiliation{Department of Physics, Montana State University, P.O. Box 173840, Bozeman, MT 59717, USA}

\author{Kelley M. Hess}
\affiliation{Department of Space, Earth and Environment, Chalmers University of Technology, Onsala Space Observatory, 43992 Onsala, Sweden}
\affiliation{ASTRON, the Netherlands Institute for Radio Astronomy, Postbus 2, 7990 AA, Dwingeloo, The Netherlands}

\author{Lucas R. Hunt}
\affiliation{National Radio Astronomy Observatory, P.O. Box O, Socorro, NM 87801, USA}

\author{Danielle Lucero}
\affiliation{Department of Physics, Virginia Tech, 850 West Campus Drive, Blacksburg, VA 24061, USA}

\author{Martin Meyer}
\affiliation{International Centre for Radio Astronomy Research (ICRAR), University of Western Australia, 35 Stirling Hwy, Crawley, WA 6009, Australia}

\author{Emmanuel Momjian}
\affiliation{National Radio Astronomy Observatory, P.O. Box O, Socorro, NM 87801, USA}

\author{Min S. Yun}
\affiliation{Department of Astronomy, University of Massachusetts, Amherst, MA 01003, USA}

\begin{abstract}

Utilizing the COSMOS HI Large Extragalactic Survey (CHILES) dataset, we investigate the evolution of the average atomic neutral hydrogen (HI) properties of galaxies over the continuous redshift range 0.09 $< z <$ 0.47. First, we introduce a simple multi-step, multi-scale imaging and continuum subtraction process that we apply to each observing session. These sessions are then averaged onto a common \textit{uv}-grid and run through a Fourier filtering artifact mitigation technique. We then demonstrate how this process results in science quality data products by comparing to the expected noise and image-cube kurtosis. This work offers the first-look description and scientific analysis after the processing of the entire CHILES database. These data are used to measure the average HI mass in four redshift bins, out to a redshift 0.47, by separately stacking blue cloud (NUV-r= -1 - 3) and red sequence (NUV-r = 3 - 6) galaxies. We find little-to-no change in gas fraction for the total ensemble of blue galaxies and make no detection for red galaxies. Additionally, we split up our sample of blue galaxies into an intermediate stellar mass bin (M$_{*} = 10^{9-10} M_{\odot}$) and a high stellar mass bin (M$_{*} = 10^{10-12.5} M_{\odot}$). We find that in the high mass bin galaxies are becoming increasingly HI poor with decreasing redshift, while the intermediate mass galaxies maintain a constant HI gas mass. We place these results in the context of the star-forming main sequence of galaxies and hypothesize about the different mechanisms responsible for their different evolutionary tracks.

\end{abstract}

\keywords{galaxies: evolution -- galaxies: ISM -- cosmology: large-scale structure of universe}

\section{Introduction}\label{sec:intro}

\quad The observation and characterization of the atomic neutral hydrogen (HI) content  of galaxies as a function of redshift is crucial for the development of a fully predictive theoretical model of galaxy evolution. Neutral hydrogen has been studied extensively in the local universe in a range of galaxy types and environments. In clusters of galaxies, the densest environments, spiral galaxies have been shown to be HI deficient \citep{haynes86,solanes01,cortese11}, and have compressed HI morphologies, or extraplanar HI gas, as a result of ram-pressure stripping from the Intra-Cluster Medium \citep{gunn72,kenney04,chung09,poggianti17}. In galaxy groups, those collections of galaxies that are less numerous and sometimes less dense than clusters, the neutral gas content of galaxies can be affected by interactions between group members and processing by the inter-group medium \citep{kilborn09,hess13}. In low density regions, or cosmic voids, galaxies are shown to have HI morphologies consistent with local optical-HI size scaling relationships for spiral and dwarf galaxies \citep{kreckel12}. Additionally, the HI phase has been characterized for massive elliptical galaxies \citep{huchtmeier94,morganti06}, as well as a range of dwarf galaxy morphologies and stages of interaction \citep{swaters02,swaters09,pearson16}.

\quad Despite the wealth of HI observations for galaxies in different environments and different morphological types in the local universe, it has been extremely challenging to directly measure HI emission in individual galaxies beyond $z$ $>$ 0.1. Some work has been done beyond this redshift, for example, the HI detection of a cluster galaxy at $z$=0.18 in \citet{zwaan01}, and the  HIGHz Arecibo survey which measured massive galaxies in the redshift range 0.17 $< z <$ 0.25 \citep{catinella08,catinella15}. The advent of broad band correlator backends made it possible to cover instantaneously a large frequency range. The new opportunity to probe a large volume at once made it worthwhile to do the deep integrations required to detect galaxies at higher redshift. The first project to do a blind survey of a large volume, probing two clusters and the large scale structure in which they are embedded in, was Blind Ultra Deep HI Extragalactic Survey \citep[BUDHIES,][]{verheijen07, gogate20}, where they sought to detect HI in individual galaxies around previously known clusters. Using the upgraded Giant Metrewave Radio Telescope (uGMRT), there have been detections of stacked HI emission at $z \approx$ 0.35 \citep{bera19}, and $z =$ 0.7 - 1.45 \citep{chowdhury20,chowdhury21}, and characterizations of the M$_{HI}$ - M$_{*}$ scaling relationship at $z \approx$ 0.35 \citep{bera23a}. These measurements have indicated very little evolution in HI gas content at $z \approx$ 0.35, but have found a significantly increased amount of HI at $z =$ 0.7 - 1.45. In contrast, the MIGHTEE-HI project \citep{maddox21}, making use of the MeerKAT radio telescope found evolution of the M$_{HI}$ - M$_{*}$ scaling relationship at $z \approx$ 0.35 \citet{sinigaglia23} and in followup studies that make use of a combined MIGHTEE and CHILES data-cube \citep{bianchetti25arxiv}. These measurements represent the current frontier of understanding for the redshift evolution of the HI content of galaxies.

\quad Further observations of neutral hydrogen across cosmic time are critical in understanding the evolution of galaxies in the Universe. Observationally, we know that the star formation rate density has decreased by a factor three from a redshift z = 0.5 to the present day \citep{hopkins06, MadauDickinson}, but we don’t know what causes the decline, as we know very little about the evolution of the gas, which is the fuel for star formation. Theoretically, it has been shown that the dominant mechanism by which galaxies gain their gas, hot vs. cold mode accretion, depends on both local environment and redshift with cold mode accretion most likely still observable at $z$ = 0 in small galaxies in voids \citep{keres05,keres09}.

\quad The COSMOS HI Large Extragalactic Survey\footnote{VLA project 13B-266.} (CHILES) is a 1000 hour integration on a single pointing in the COSMOS field with the Karl G. Jansky Very Large Array\footnote{The National Radio Astronomy Observatory is a facility of the National Science Foundation operated under cooperative agreement by Associated Universities, Inc.} (VLA) in the B configuration. CHILES observes a frequency range of 960 to 1420 MHz, allowing for the direct detection of galaxies with an M$_{HI}$ of 3 $\times$ 10$^{10}$ M$_{\odot}$, out to a redshift of 0.47. The first 178 hours of CHILES data have been successfully used to perform several different science verification studies. \citet{fernandez16} present the highest redshift direct detection of HI in emission at $z$ = 0.376 and find a galaxy with an unusually high star-formation rate for its stellar mass. Note, that \citet{heywood24} obtained a very low upper limit for the same galaxy, which will be discussed in Blue  Bird et al (in prep). \citet{hess19} demonstrate that even in a frequency range severely affected by radio frequency interference (RFI) there are stretches where galaxies can be detected with reliable extended morphologies. \citet{bluebird20} investigate the spin alignment of nearby galaxies with respect to the orientation of the cosmic web filaments, as traced by the Discrete Persistent Source Extractor \citep[DisPerSE,][]{sousbie11} algorithm and customized by \citet{luber19} for the CHILES survey. In addition, in \citet{dodson22} we developed a successful imaging routine using the first 178 hours of CHILES data that utilizes an LST-dependent \textit{uv}-based modelling and subtraction for sources that are beyond the half-power point of the VLA primary beam. This procedure properly accounts for the hour-angle dependence of the VLA primary beam and allows for the most accurate modeling of the flux density of sources beyond the half-power point. In order for this routine to use this highly precise approach  large computational resources, such as those available through Amazon Web Services (AWS), and the computing framework introduced in \citet{wu17}, are required \citep{dodson16}.  

\quad The calibration and imaging of the CHILES data are complicated by the large volume (approximately 220 TB of raw data), and long duration (six years\footnote{Beginning in November 2013 and ending in April 2019}) while requiring regular checks on the quality of the calibration and imaging. Additionally, our observations extend beyond the protected bands for radio astronomy resulting in some of the bandwidth being significantly affected by RFI. As more data were calibrated and we integrated deeper, low level image artifacts, such as residual sidelobes from a source three degrees from the field center, were found to be affecting the science field. As a result, it became apparent that a rapid turn around to assess data and image quality was necessary in order to make informed decisions about how to proceed with data processing. In order to address theses issues, here we introduce a quick and efficient imaging routine, that produces science quality images. The technique we introduce in this work does not make all the appropriate considerations that \citet{dodson22} do, but produces data cubes with a rapid turnaround time that can be used to assess data quality, as well as for scientific pursuits, and offers a suitable alternative for imaging large interferometric databases. As such, this work does not supersede that presented in \citet{dodson22}, but offers an alternative methodology that can be used for scientific analysis while the full database subtraction using the technique in \citet{dodson22} can be completed.

\quad This paper is structured as follows: In Section \ref{sec:chiles_data}, we discuss the CHILES data, and the imaging techniques utilized to produce the cube used in our analysis, and go on to describe the ancillary data in the COSMOS field that we used in Section \ref{sec:ancillary_data}. In Section \ref{sec:stacking}, we outline the procedure for HI and continuum stacking implemented for our analysis, and in Section \ref{sec:results} summarize the results of the stacks we conduct. In Section \ref{sec:disc} we discuss our findings within the broader context of previous literature results and in Section \ref{sec:sum}, we summarize our findings. In this work, we assume a standard flat $\Lambda$CDM cosmology with H$_{0}$ = 70 km s$^{-1}$ Mpc$^{-1}$, $\Omega_{\Lambda}$ = 0.7 and $\Omega_{M}$ = 0.3.

\begin{deluxetable}{cc}
\tablecaption{Summary of Observations\label{tab:obs_sum}}
\tablecolumns{1}
\tablenum{1}
\tablewidth{0pt}
\tablehead
{
\colhead{Property} &
\colhead{Value}
}
\startdata
Target Pointing & 10h01m24s +02$^{\circ}$21$\arcmin$00$\arcsec$\tablenotemark{a} \\
Observation Dates & 2013/10/25 - 2019/04/11\tablenotemark{b} \\
Total Observing Time & 1062 hrs\tablenotemark{c,d} \\
Bandpass $\&$ Flux Density Scale & 3C286 \\
Complex Gain Calibrator &  J0943–0819 \\
Correlator Integration Time & 8s \\
Spectral Window Setup & 15 x 32MHz \\
Channel Width & 15.625 kHz \\
Continuous Frequency Coverage & 960 - 1420 MHz \\
Synthesized Beam & 5$\farcs$5 - 8$\farcs$5\tablenotemark{e} \\
\enddata
\tablenotetext{a}{In the J2000 coordinate standard.}
\tablenotetext{b}{The observations were taken over five different epochs, corresponding to five consecutive VLA-B configurations. This results in the observations being taken over different parts of the year and having a mix of both day and night time observations.}
\tablenotetext{c}{The total integration time is split amongst the five observing epochs with 188, 220, 190, 237, 192 hours, respectively.}
\tablenotetext{d}{Individual sessions have total observation times ranging from 1 - 8 hours, with a mean time of 4.75 hours, and approximately 45\% of observations being 6 hours.}
\tablenotetext{e}{The range of synthesized beam major axis length. The size increases within this range with decreasing frequency.}
\end{deluxetable}

\section{CHILES Data and Imaging}\label{sec:chiles_data}

\subsection{The CHILES Database}

\quad The CHILES observations were taken with the VLA in its second most extended configuration (B). This configuration has a maximum baseline length of $\approx$11km, corresponding to a synthesized beam of major axis 5$\farcs$5 to 8$\farcs$5 at 1420 and 960 MHz, respectively, which in astrophysical terms corresponds to a physical resolution of 0.6 - 47.8 kpc for redshifts $z$ = 0.005 - 0.48. Assuming a velocity width of 150 km s$^{-1}$, our 3$\sigma$ HI mass detection limit at $z$ = 0.48 is 3.0$\times$10$^{10}$ M$_{\odot}$. This allows us to simultaneously have detailed HI maps at the lowest redshifts, and be able to identify individual galaxies at the highest redshifts. The observations use 60 of the 64 available correlator baseline board pairs to achieve the desired 30,720 channels spread across 15 spectral windows each of 32 MHz of bandwidth, in each observing session. To minimize loss of sensitivity at the edges of the band of each spectral window, we used frequency dithering. Specifically, we shifted the edge of the lowest frequency in steps of 5 MHz, and used three different settings per epoch, and different settings across the five epochs. This resulted in almost uniform sensitivity across the 460 MHz band. A summary of the properties of the observations, including the pointing, observing dates, instrumental setup and calibrators, can be found in Table \ref{tab:obs_sum}. 

\quad The frequency range covered by CHILES is impacted by a variety of RFI sources of varying magnitudes occurring at different frequencies. The lowest frequencies (approximately 960-1080 MHz) remain relatively RFI free, with the only exception being a narrow spike of RFI at 1030 MHz due to aeronautical ground-to-air radar. Similarly, the highest frequencies (approximately 1300-1420 MHz) are relatively RFI free with exception of several narrow spikes due to GPS satellites and aeronautical radar. The frequency range in the middle (1080-1300 MHz) is the most severely affected range with several different sources of RFI. This includes RFI from aeronautical radar, GLONASS and GPS satellites, and communications from the nearby airport in Albuquerque.

\quad The data are calibrated with a modified version of NRAO’s calibration pipeline for the VLA data using CASA version 5.3.0 (Pisano et al. in prep) at the West Virginia University High Performance Computing Spruce Knob facility\footnote{Computational resources were provided by the WVU Research Computing Spruce Knob HPC cluster, which is funded in part by NSF EPS-1003907.}. In total, 856 hours have successfully been processed through the calibration pipeline with some data excluded due to offline processing issues. 

\subsection{Imaging}

\subsubsection{Continuum Subtraction}\label{sec:cont_sub}

\quad The production of a usable HI spectral line cube requires all radio continuum to be subtracted from it for HI analysis. There are three basic ways to subtract the continuum in the Common Astronomical Software Application \citep[CASA,][]{casa}: \textit{imcontsub}, \textit{uvcontsub}, and \textit{uvsub}. They are described as follows:

\begin{enumerate}
    \item \textit{uvcontsub:} This is a purely visibility domain based continuum subtraction method in which a polynomial is fit, per baseline, to the visibilities, and then subtracted off. Ideally, this method would use only the HI line-free emission channels in the fit, but this requires them to be known a priori.
    \item \textit{imcontsub:} This is a purely image domain based continuum subtraction method in which a polynomial is fit to to each pixel in the image domain and then subtracted off. Ideally, this method would use only the HI line-free emission channels in the fit, but this requires them to be known a priori.
    \item \textit{uvsub:} This method is a hybrid imaging and visibility method. In this method a continuum image is made and then the Fourier transform of the CLEAN components are subtracted from the visibilities.
\end{enumerate}

\quad Each method has benefits and limitations associated with it, and oftentimes, combinations of the methods are used to achieve optimal image cubes. For a summary of these tasks, and their strengths and weaknesses, as implemented in the Astronomical Image Processing System \citep[AIPS,][]{greisen03}, see \citet{cornwell92}. 

\quad For the CHILES data, the use of \textit{uvcontsub} is not optimal due to the fact that the implementation in CASA is not effective in removing sources far from the field center. Normally, this is not an issue, but in the CHILES data, we are able to detect sources beyond the first null of the VLA primary beam due to our excellent sensitivity. As a result, the use of \textit{uvcontsub} results in residual image plane artifacts. The use of \textit{imcontsub} is complicated by the fact that in order for it to be useful, we would need to image and CLEAN a very large image for the whole database which is extremely computationally expensive. Additionally, the sources are affected by a standing wave, with an approximately 17 MHz period, due to a central blockage affecting aperture efficiency \citep{jagannathan18}. In order to account for this we would need to use an extremely high order polynomial which could very well subtract real emission. Additionally both \textit{imcontsub} and \textit{uvcontsub} could slightly bias our HI source parameters, as this is a blind survey, we do not a priori know where the HI emission is and include it in spectral fits. For the stronger sources we can correct for this later by subtracting a fit that excludes the HI channels.

\quad This leaves us with the use of \textit{uvsub} which we employ at two different spatial resolutions, and in three different steps. Furthermore, rather than commit large amounts of computing resources to this effort, we grid our data, in \textit{uv} space, and average each session on to a common grid, using the CASA task \textit{msuvbin}, described in \citet{memo198}. This allows us to reduce our entire 220 session database to one common \textit{uv} grid that can then be used in the subsequent final HI cube creation. Below, we present the three step radio continuum subtraction, entirely contained within CASA 5.6, using three iterations of \textit{uvsub} and an iteration of \textit{imcontsub}. The radio continuum subtraction and data combination method we implement for the CHILES data is done on a per-session basis for the continuum subtraction, and then combined in the \textit{uv}-plane. It is performed in the following manner:

\begin{enumerate}
    
    \item We split off the target data from a full single session \textit{uv} dataset, and spectrally average the data by a factor 4, to a resolution of 62.5 kHz (13.2 km s$^{-1}$ at $z$ = 0). This channel width is chosen because it is a compromise between the required sensitivity for HI detection, and the velocity resolution required to properly deduce kinematics of our HI detected galaxies.
    
    \item We produce a cube spanning the entire bandwidth, with 2 MHz channels to properly sample the standing wave spectral response, for the inner 1.14$^\circ$ of the field, at the native angular resolution, approximately 5$\arcsec$. For this image, we use a robust value of 0.5, as implemented in the tclean task of CASA, and 512 w-projection planes to correct for the effects of a non-coplanar array and sky curvature for the production of widefield image \citep{cornwell05}. We implement a targeted CLEAN, using a mask over known continuum sources that was created using a continuum image generated by the CHILES data. We CLEAN the sources within the mask down to 400 $\mu$Jy, corresponding to the approximately 1$\sigma$ level for a four hour session. The Fourier transform of the CLEAN components are then subtracted from the \textit{uv} data using \textit{uvsub}. This subtracts all of the in-field continuum sources.
    
    \item We produce a cube spanning the entire bandwidth, with 2 MHz channels to properly sample the standing wave spectral response, but this time for a field that of diameter 2.84$^{\circ}$, and using a Gaussian \textit{uv}-taper that creates a resolution element of 25$\arcsec$. For this tapered image, we use 1024 w-projection planes.We CLEAN the sources within the mask down to 300 $\mu$Jy, corresponding to the approximately 1$\sigma$ level for a four hour session at this tapered resolution. The Fourier transform of the CLEAN components are then subtracted from the \textit{uv} data using \textit{uvsub}. This subtracts the strong out-of-field continuum sources.
    
    \item We, for the last time, create an image cube with a 2 MHz channel width across the entire bandwidth, over a small region, approximately 8.6$\arcmin$, covering a strong source, 4C-00.37, approximately 3$^{\circ}$ south of our field with an observed flux of 1.5 mJy at our lowest frequencies, at the native resolution. The data are phase-shifted to the position of this problematic source, and for this small image, we use no w-projection planes and robust value of 0.5. We provide a CLEAN mask over the known continuum source and CLEAN down to the 400 $\mu$Jy, corresponding to the approximately 1$\sigma$ level, within this mask. The Fourier transform of the CLEAN components are then subtracted from the \textit{uv} data using \textit{uvsub}. This subtracts the last remaining detectable continuum source in our field.
    
    \item We perform a final round of RFI flagging using the statistical flagging task in CASA, \textit{rflag}, at the 5$\sigma$ level on the fully uvsubbed data. We then average the data for all databases on a common \textit{uv}-grid, performing an on-the-fly Doppler correction before the combination, at the desired channel width, and image size, using the CASA task \textit{msuvbin}, in the LSRK velocity reference frame. We leave out data with $|u|$ $<$ 50m to eliminate horizontal stripes in the image caused by RFI and imaging artifacts from the projected baselines distributions. This \textit{uv}-gridded database is then imaged with \textit{tclean} with the specified pixel and image size and from the \textit{msuvbin} step, and any other required imaging parameters, such as \textit{uv}-tapering and spectral averaging. This step produces an image cube that spans the entire frequency bandwidth with the required specifics.
    
    \item We run an implementation of \textit{imcontsub}, with a linear fit, that is run over the entire bandwidth of the cube. After this, we do any image post-processing necessary, such as that in Section \ref{sec:uvflag} in order to produce a final data product.
    
\end{enumerate}

\quad Steps one through four are done with a single script and, on average, have a wall time, the elapsed time taken by computing, of 20 hours and require 58 GB of memory. Using the WVU HPC Thorny Flat\footnote{Computational resources were provided by the WVU Research Computing Thorny Flat HPC cluster, which is funded in part by NSF OAC-1726534.}, we can run 10 sessions simultaneously, which translates into a total wall time of 18.5 days to get all 222 sessions through steps one to four. Step five takes approximately one hour for every session that is being averaged in the \textit{uv}-plane, which for the CHILES data translates to approximately 9.5 days, and requires, on average, 256 GB of memory. Step six is completed in approximately 12 hours, and utilizes 48 GB of memory. Therefore, from beginning to completion, for the entire CHILES database, this procedure takes 28.5 days, and requires resources ranging from 48 GB of memory to 256 GB of memory. The final data product, an image cube of 1200 by 1200 pixels and 3680 channels, is approximately 20 GB in size.

\begin{figure*}
\begin{center}
\includegraphics[width=\textwidth]{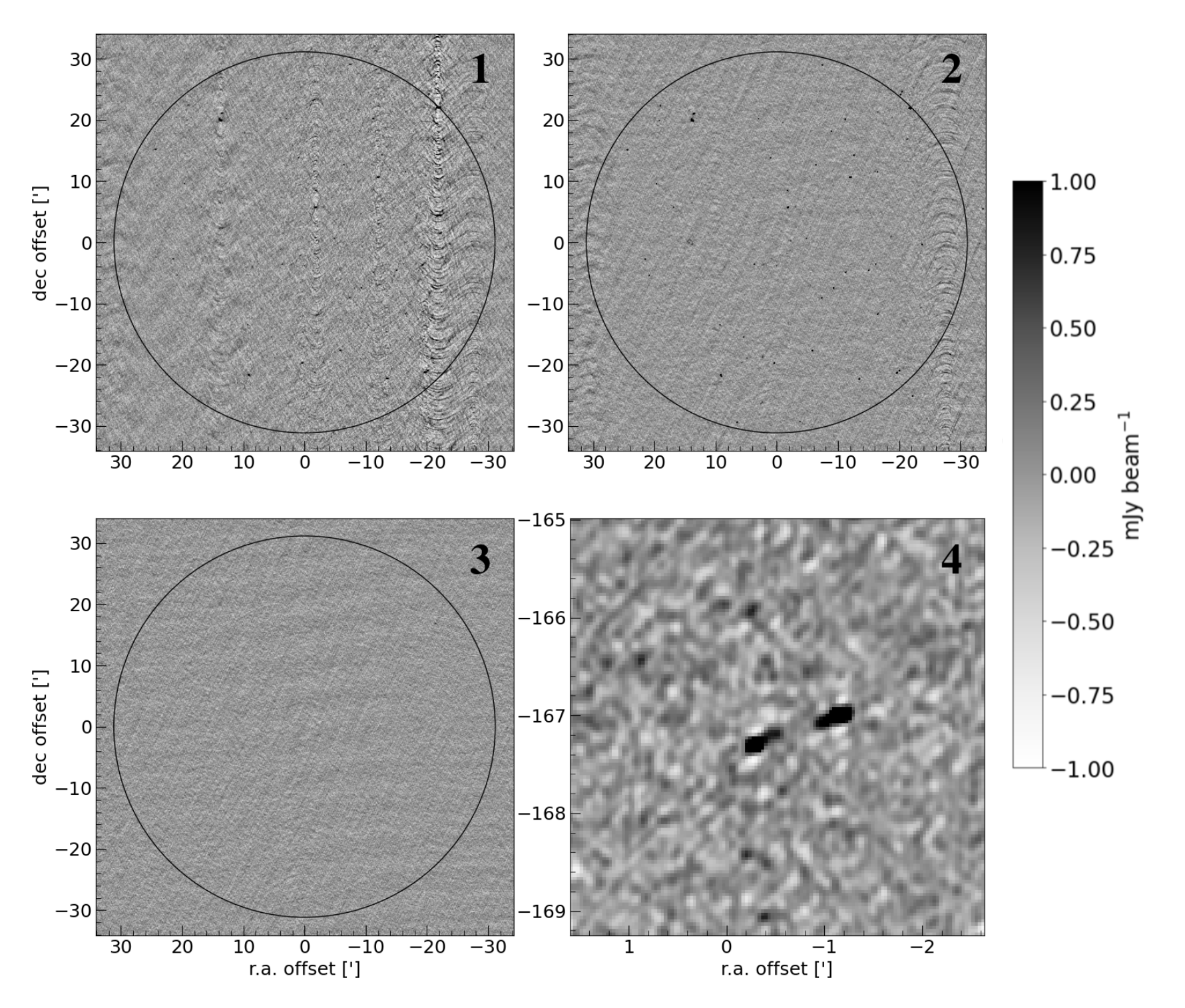}
\caption{An example of the continuum subtraction routine on a single six hour observation run, for a single channel of width 2 MHz, centered on 961 MHz. In panel 1, we show an unCLEANed image of the science field, in panel 2-4, we show the CLEANed pre-subtraction images produced in each of those steps, whose fourier components are then subtracted from the visibilities using \textit{uvsub}. The black circle corresponds to the 20\% point of the primary beam of the VLA at these frequencies, and the grey scale is common for all images, and illustrated by the grey bar to the right. For each image the x and y axis are defined by the angular offset from the CHILES field phasecenter. This changes for some images as images 1,2,3 are of field center, while image 4 is centered on a problematic source, and images 1,2,4 have 2$\arcsec$ pixels and image 3 has 5$\arcsec$ pixels.}
\label{fig:imaging_workflow_example}
\end{center}
\end{figure*}

\begin{figure*}
\begin{center}
\includegraphics[width=\textwidth]{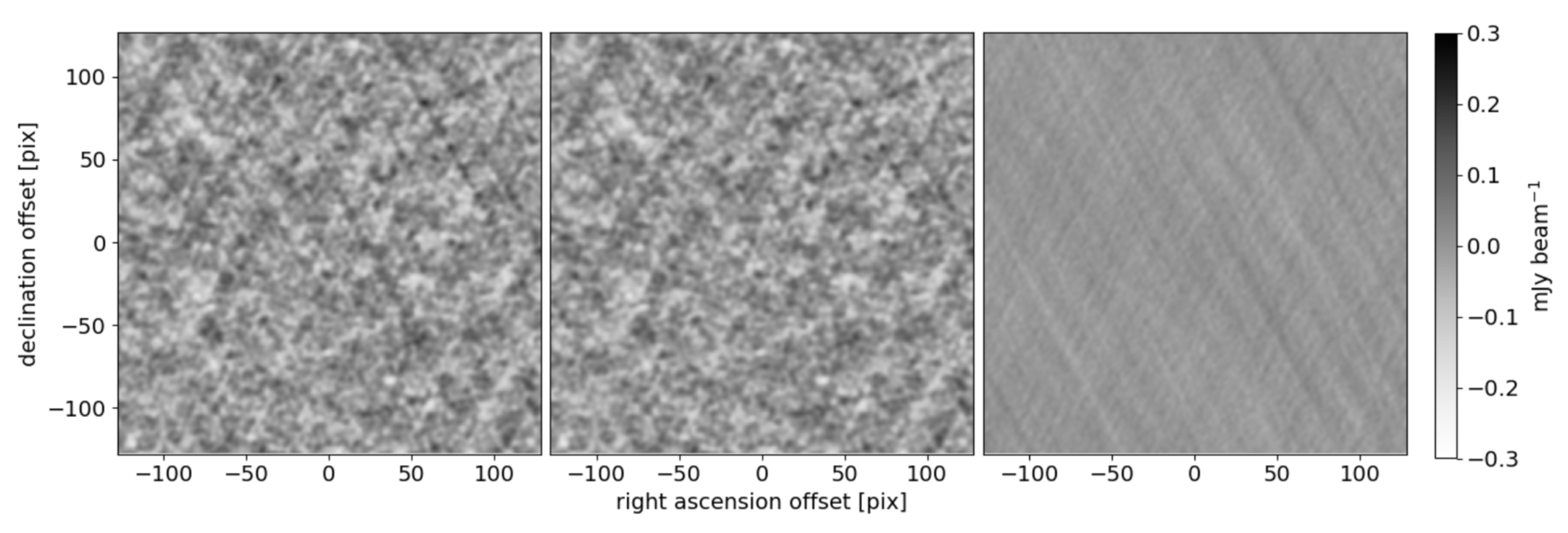}
\caption{\textbf{\textit{Left:}} A single channel input, at 1153.25 MHz, before the flagging routine, with a r.m.s. noise of 51.5 $\mu$Jy beam$^{-1}$. \textbf{\textit{Middle:}} The same data after the flagging routine has been run, with a r.m.s. noise of 49.3 $\mu$Jy beam$^{-1}$. \textbf{\textit{Right:}} The difference between the images before and after the flagging routine. Each image has a common grey scale, ranging from -0.3 to 0.3 mJy.}
\label{fig:uvbinflag_chan}
\end{center}
\end{figure*}

\quad In Figure \ref{fig:imaging_workflow_example}, we show examples of steps one through four for a single 2 MHz channel, at 960MHz, for a single six hour observing session. In the first panel, we show an image that represents a simple FFT of the data with no deconvolution or continuum subtraction. This image shows significant side lobes from sources both within, and outside of, the 20\% point of the primary beam, indicated by the black circle. In the second panel, we show the CLEANed image described in Step 2, in which the in-field sources from this image are subtracted. The sidelobes to the extreme east and extreme west of the image are still present as the sources they come from are outside this field. In panel three, we show an image after the CLEANed image described in Step 3, where the sources beyond the 20\% point of the primary beam have been removed with \textit{uvsub}. The image created in this step is in actuality four times the size of the image in this panel, but we display a zoomed in image to just encompass the science field, the inner quarter of the image produced in Step 3, to show that the sidelobes have been removed, and additionally, a low-level north-south ripple across the center of the image caused by the source 3$^{\circ}$ to the south. Lastly, in panel four, we show a pre-subtraction image from this observing run of the source that is subtracted in step four. This image shows how, quite remarkably, a source 3$^{\circ}$ away from field center can still be detected, and thus causes sidelobes that affect the science field. As this panel shows, this source is able to be successfully modelled, CLEANed, and subsequently subtracted thus successfully mitigating any systematic offsets from the resulting residual sidelobes.

\quad These panels illustrate how each step is neccessary in the reduction of the prevalence of imaging artifacts. An important reminder is that this is just a single session, the full CHILES database is more than a factor 10 times more sensitive than this session. This means that any imaging artifacts seen in these images will have appeared at a much more significant level using the full database, had we not corrected for them with this routine. Additionally, in the above points we determine the CLEAN threshold based upon the number of visibilities in a 4 hour session, the average length of a CHILES session. We use this same threshold for all databases, independent of total time, as observations are never more than a factor two shorter or longer than four hours. For shorter observations, this may cause an increase in run time due to the lower threshold, in terms of signal-to-noise ratio, and for the longer observations, it may cause some residual continuum to remain in the image. Only the latter will affect the image quality and is mitigated by the last image-plane subtraction in step six.

\begin{figure*}
\begin{center}
\includegraphics[width=0.98\textwidth]{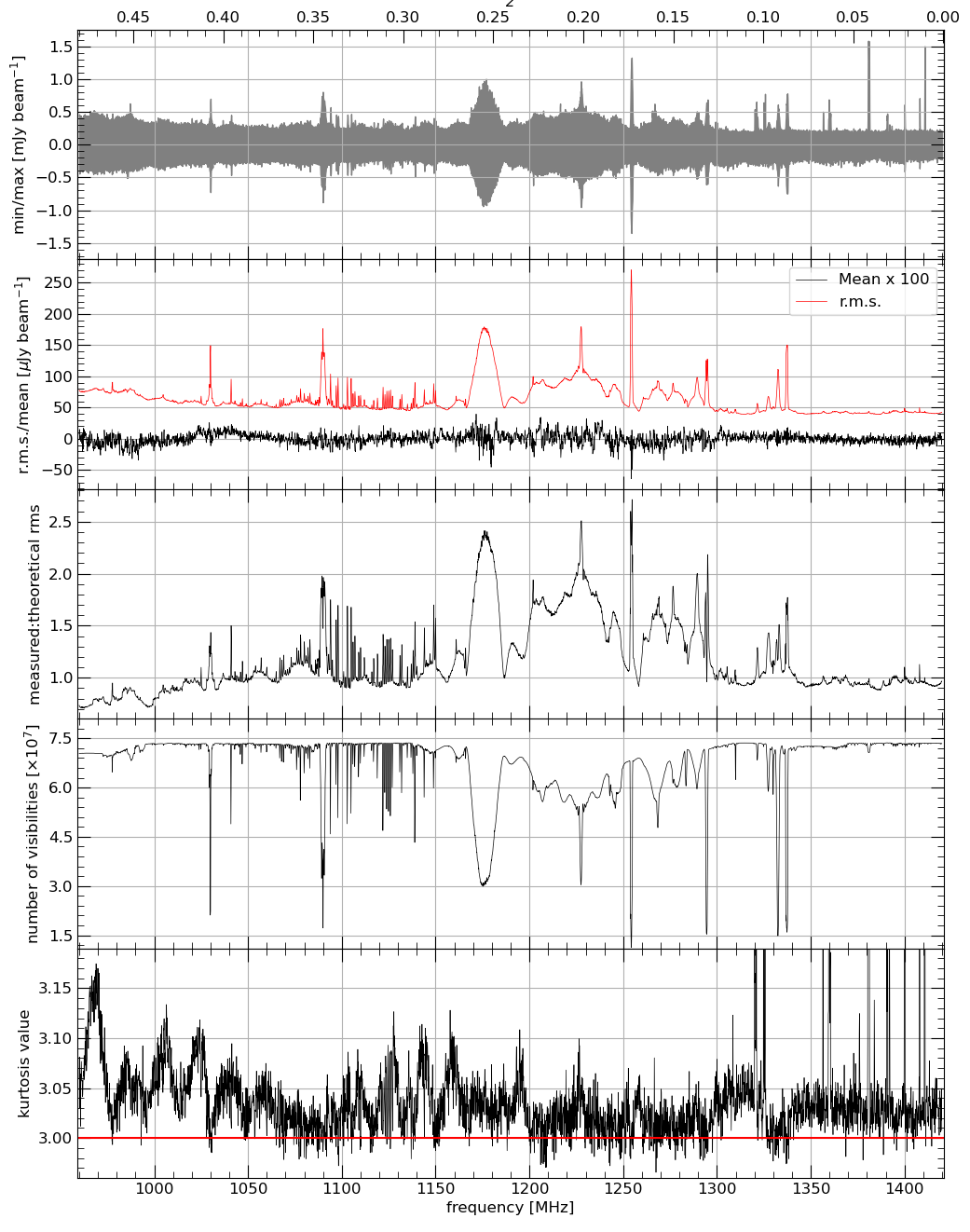}
\caption{As a function of frequency (enumerated on the bottom axis with corresponding redshifts enumerated on the top), we show the \textit{(top row)} range of all pixel values, the \textit{(second row)} r.m.s. noise and mean pixel value multiplied by 100, the \textit{(third row)} ratio of measured noise to theoretical noise, the \textit{(fourth row)} number of unflagged visibilities used in imaging, and the \textit{(bottom row)} kurtosis values of the CHILES data, following the processing outlined in Section \ref{sec:chiles_data}}
\label{fig:QA}
\end{center}
\end{figure*}

\subsubsection{Additional RFI Mitigation}\label{sec:uvflag}

\quad In the spectral line cubes produced after our imaging pipeline, there are still some lingering effects of RFI and imperfect calibration. Both of these effects are seen in the image-domain as broad stripe-like features that stretch across the images. As a result of their similarity, and the convenience of Fourier-transforms, we approach these problems with a \textit{uv}-plane mitigation technique. Any feature that appears with broad image-wide characteristics will appear as localized features in a Fourier transform of the image. The \textit{uv}-space RFI mitigation technique we use is an implementation of the method introduced in VLA Memo 198 \citep{memo198}, and further explored in \citet{sekhar18}, optimized for use with the CHILES data. We use this adaptation on the spectral line data produced following the completion of the six steps outlined in Section \ref{sec:cont_sub}. The adaptation we use is completely native to Python, only requires the NumPy package, is done on a per-channel basis, and is described as follows:

\begin{enumerate}
    \item Take a Fast-Fourier Transform (FFT) of the image plane data.
    \item Rotate each quadrant in \textit{uv} space such that \textit{u} and \textit{v} = 0 are in the lower left corner, and average the four quadrants. There is some redundancy as quadrants I and III, and II and IV are identical, but this does not affect the results, as the threshold is chosen after the averaging and optimized to remove artifacts and conserve source flux. This optimization is verified by investigating a number of thresholds and selecting the lowest threshold that has no effect on the flux of our brightest and most extended source.
    \item Bin the data by baseline length in increments of ten meters, calculate the mean and standard deviation in each bin, and if the value in a cell is greater than 2$\sigma$ above the mean for the radial profile at the corresponding baseline length of the cell, set the point equal to zero.
    \item FFT the \textit{uv}-space data back into the image-domain.
\end{enumerate}

\quad This technique has been applied to the final cube, representing the aggregate of all CHILES data, whose production is described in Section \ref{sec:cont_sub}. The technique was parallelized on a per-channel basis, and can be run on the WVU HPC, Thorny Flat, in approximately four hours, requiring 48 GB of memory, for a 20 GB image cube of 3680 channels and 1200 by 1200 spatial pixels. In Figure \ref{fig:uvbinflag_chan}, we show the effect of this uvbin flagging method on the CHILES data. The right-most panel of this figure shows the residual features that this technique successfully eliminates from the image. The residual features in this image are various diagonal stripes running across the image. The diagonal stripes are likely caused by a combination of two reasons: low-level RFI that the automated flagging missed as part of the calibration pipeline, and artifacts from the imperfect removal of sources at, and beyond, the half-power point of the VLA primary beam. The removal of both of these effects is expected, as they can both be similarly characterized as large-scale variations across the image. As a result, they appear similarly in the \textit{uv}-plane and are successfully excised by our technique.

\begin{figure*}
\begin{center}
\includegraphics[width=\textwidth]{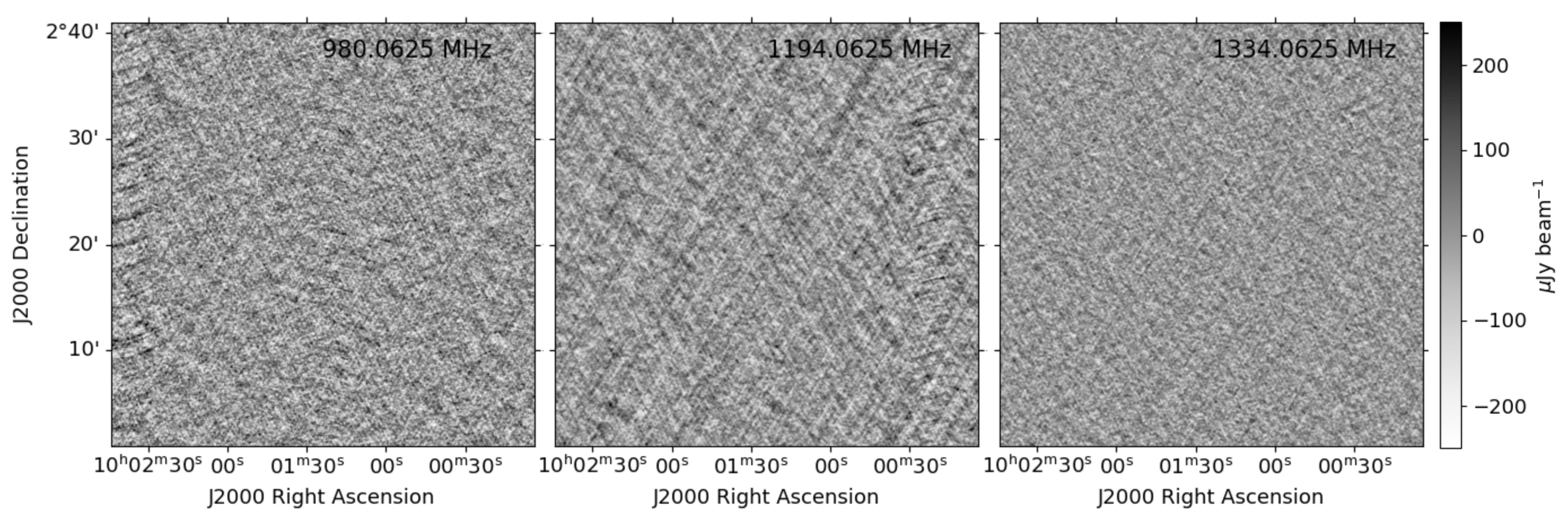}
\caption{Three channel maps located at 980.0625 (left), 1194.0625 MHz (middle), and 1334.0625 MHz (right) pulled from the final cube used in our analysis. Each image has a common grey scale, ranging from -0.25 to 0.25 mJy. These represent exemplary channels of intermediate quality (left), poor quality, less than 10\% of the data, (middle), and exceptional quality (right).}
\label{fig:ExampleChannel}
\end{center}
\end{figure*}

\begin{figure*}
\begin{center}
\includegraphics[width=0.75\textwidth]{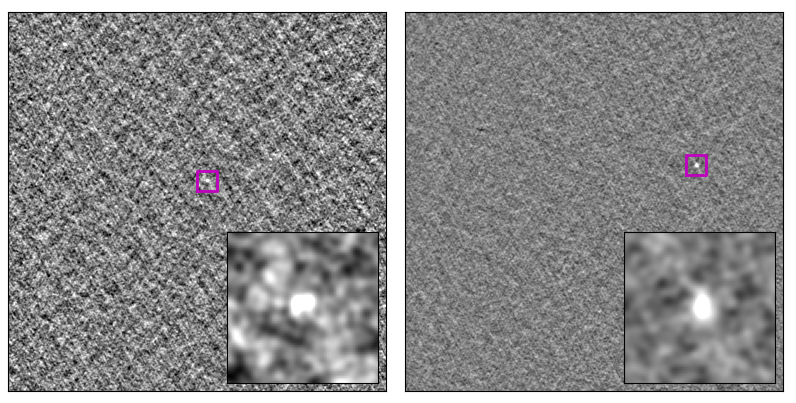}
\caption{An example of two channels that contain emission from a previously published CHILES HI detection. Each panel represents the entire CHILES field of view, and the zoom-in on the bottom right corners are the 2' region around the detection, corresponding to the magenta boxes. Each image has a common grey scale, ranging from -0.25 to 0.25 mJy. The  \textit{Left:} This is a detection at 1216.25 MHz, originally presented in \citet{hess19} using 178 hours of the CHILES data. \textit{Right:} This is a detection at 1356.5 MHz, originally presented in \citet{bluebird20}, using 178 hours of the CHILES data.}
\label{fig:ExampleDetection}
\end{center}
\end{figure*}

\quad The r.m.s. noise in the pre-flagging image is 51.5 $\mu$Jy beam$^{-1}$ and is slightly reduced in the post-flagging image to 49.3 $\mu$Jy beam$^{-1}$. The artifacts visible in the difference image that were successfully removed lie at, or below, the 1$\sigma$ level, and in the two images from pre and post the technique it is hard to detect the effect of the technique by eye. However, it is crucial to remove these artifacts as they can have a profound impact on source structure and characteristics. For example, a 0.5$\sigma$ ripple would introduce a $\pm$16.7\% error in flux density on a 3$\sigma$ detection, if the detection lies on a peak or trough of the ripple.

\subsection{Final Data Quality} \label{sec:QA}

\quad In order to validate the output of the procedure outlined in Sections \ref{sec:cont_sub} and \ref{sec:uvflag}, we can compare the empirical results to that of expectations. If they are similar, it is an indication that the continuum subtraction has been successful. In Figure \ref{fig:QA}, we show the range of all pixel values (top) and the ratio of measured to theoretical noise (third row) for all channels. The theoretical noise is calculated using the frequency dependent VLA System Equivalent Flux Density (SEFD), and the number of unflagged visibilities used in the cube creation. Strictly speaking this calculation is only correct for an image-cube created with naturally weighted visibilities, while our cube is made with a briggs robust value of 0.5. However, our choice of a briggs robust parameter has a minimal, $<\approx5\%$, effect on the measured noise in an image-cube \citep{briggs95}. Interestingly, in the third panel the ratio occasionally dips to slightly below 1. We theorize that this is due to the fact that the SEFD used in the theoretical calculation is not a 100\% accurate measurement. This is because there is some natural variation of SEFD from antenna to antenna. As a result, we only consider a deviation in the ratio of 30\% or greater as significant, as any deviation lower than that could be a result of these somewhat inaccurate use of SEFD measurements. There is some deviation greater than this threshold at our lowest frequencies, but we believe this is due to the fact that these frequencies represent the extreme low end of the VLA L-band receivers where the SEFD is least well-behaved.

\quad Simply having noise that is comparable to theoretical, however, is not the only necessary indicator for science-ready data. An outstanding issue for deep radio interferometric surveys is the presence of non-Gaussian noise. Non-Gaussian noise is caused by calibration/deconvolution errors, RFI, and subsequent incomplete/incorrect continuum subtraction which can significantly affect the scientific results of the data. To inspect this issue in the CHILES data, we show the kurtosis value (the fourth moment of a Gaussian distribution) as a function of frequency in the bottom row of Figure \ref{fig:QA}. The horizontal red line at a kurtosis value of three in the bottom row of Figure \ref{fig:QA} indicates the expectation value for perfectly Gaussian data, with the expected variation on three, for CHILES-like data, being 0.004. Values of kurtosis that are greater than or less than three indicate that the distribution has more or fewer extreme values than expected, respectively \citep{balanda88}. The expectation for this final CHILES image-cube, with perfect calibration and RFI excision, should be that the visibilities have a Gaussian distribution, except for the regions for which their is HI emission, and thus a perfectly gaussian image in emission-free channels. The sharp spike features seen in the bottom panel of Figure \ref{fig:QA} above 1300 MHz can be attributed to known bright HI detections. The flux from the HI emission, as well as the bright sidelobe patterns, injects those image planes with pixel values that significantly deviate from Gaussian noise, and present as sharp image-plane features.

\quad Taken in its totality, Figure \ref{fig:QA} offers some insight into the complicated nature of the CHILES data. The kurtosis values indicate that the noise in the data departs from Gaussianity, in several frequency ranges, most notably in 960-1060, 1100-1200, and 1300-1320 MHz, with the largest deviations at the lowest frequencies. However, the noise ratio only significantly increases in the range 1150-1300 MHz. From these two facts, we can surmise that the data in the frequency range 1150-1300 MHz suffers strongly from residual RFI and imaging artifacts, but that the entire bandwidth suffers from non-Gaussian noise on a lower level. This second point is key in understanding this dataset as it emphasizes the fact that we must treat each HI detection and HI stacking result with scrutiny to ensure it is not an imaging artifact, or a combination of imaging artifacts, respectively. 

\quad The statistical overview presented by the panels in Figure \ref{fig:QA} offer a comprehensive overview of the data, and in Figure \ref{fig:ExampleChannel} we complement this by presenting three channel maps from our final cube. Specifically, in Figure \ref{fig:ExampleChannel} we present three channel maps at three different representative frequencies. We show the data at 980.0625 MHz, which is an example of a channel that still has some continuum subtraction residuals on high right ascensions, 1194.0625 MHz, which is an example of a channel that has noticeable RFI and continuum residuals across the entire image, and 1334.0625 MHz, which shows an ideal channel with very little artifacts. A significant potion of our data, especially at frequencies above 1300 MHz appear as immaculate as our example channel at 1334.0625 MHz. We estimate that in the range 960 to 1300 MHz no more than 10\% of the channels are as bad as the channel shown at 1194.0625 MHz and thus not suitable for science, with that example channel being amongst the most extreme examples of a non-ideal channel. Note, that even in the worst range, 1150 to 1300 MHz, useful scientific results can be obtained \citep[][Hess et al. in prep]{hess19}. The channel maps, in general, show data of the type of quality that would be suggested by the plots in Figure \ref{fig:QA} at the specific frequencies. In Figure \ref{fig:ExampleDetection}, we show an example of two detections in our 856 hour cube. In the left panel, we see an example of a detection at 1216 MHz which is in the range most heavily contaminated by RFI, and in the right panel, we show an example a channel in our most pristine frequency range. These images demonstrate the suitability of our final image for detecting HI, even in frequency ranges most affected by RFI.

\quad The above mentioned imaging and RFI mitigation techniques have been applied to all successfully calibrated data (856 hrs). The data quality is slightly less than what we hoped for with a typical rms noise of 50 $\mu$Jy beam$^{-1}$ per 125 kHz channel as opposed to the proposed 50 $\mu$Jy beam$^{-1}$ per 31.25 kHz channel. This and its consequences will be discussed in a forthcoming paper (van Gorkom et al. in prep). However, our results in Figure \ref{fig:QA}, demonstrate that we achieve reasonable values of measured noise based on the actual numbers of visibilities used in our cube production.

\section{Ancillary Data}\label{sec:ancillary_data}

\quad Throughout this work we make use of the extensive multi-wavelength data, and subsequent derived galaxy properties, made publicly available by the COSMOS collaboration \citep{scoville07}. From the public COSMOS data, we use the COSMOS2008 ID as well as the rest frame NUV-r color, and derived stellar masses, calculated with the stellar population model of \citet{bruzal03} and the initial mass function of \citet{chabrier03}, as presented in \citet{laigle16}. In addition to the COSMOS data, we also utilize data from the Galaxy and Mass Assembly (GAMA) survey \citep{driver11}. Specifically, we use the GAMA collaboration's reprocessing of COSMOS multi-wavelength data in order to derive the highest possible confidence redshifts \citep{davies15}. For the HI stacking done in this work it is critical to have a-priori redshifts with errors similar or less than the typical HI line width in the stack, in order to ensure that we are able to correctly identify the expected line center of the HI detection, and not smear the average spectrum \citep{maddox13}.  As a result, in all input galaxies for the HI stacks in this paper we only use galaxies with the highest confidence redshifts, as specified in \citet{davies15}. 

\quad In addition to publicly available COSMOS data, we make use of the radio continuum data from the CHILES Continuum/Polarisation Survey (CHILES Con Pol). CHILES Con Pol is a 1000 hour 1.4 GHz wideband full polarization radio continuum deep field that was observed commensurate with the CHILES HI deep field, utilizing the four baseline board pairs of the VLA correlator that were not used by CHILES. Further details on the calibration, imaging, and data quality can be found in \citet{memo208} and Luber et al. (submitted). We utilize this data in order to estimate star-formation rates, in the methodology described below in Section \ref{sec:cont_methodology}.

\section{Stacking Methodology} \label{sec:stacking}

\subsection{Neutral Hydrogen Emission} \label{sec:HI_methodology}

\quad In order to produce an accurate measurement of the average neutral hydrogen content of a sample of galaxies, we perform a cubelet stack with the following specifications (for a detailed description of cubelet stacking see \citet{chen21a} and for a demonstration of its capabilities see \citet{chen21b}):

\begin{enumerate}

    \item Extract a cubelet centered in position and frequency on a source. We chose to extract cubelets with 64 pixels in right ascension and declination, and 89 frequency channels, corresponding to a cube with dimensions 128$\arcsec$ x 128$\arcsec$ x 11.125 MHz. At the lowest and highest redshifts we use in our stack, 0.09 and 0.47, this corresponds to physical dimensions of 236 kpc x 236 kpc x 2565 km s$^{-1}$ and 764 kpc x 764 kpc x 3474 km s$^{-1}$, respectively.
    
    \item Perform a test on the cubelet to see if the noise is sufficiently Gaussian. We define this as having no more than three times the expected number of pixels greater than $\pm$3$\sigma$ for a normal distribution. Cubelets that do not meet this criterion are not used in the subsequent stacking. These rejected cubelets account for approximately 15\% of the total number of galaxies within the sample.
    
    \item Perform the appropriate primary beam correction by multiplying the entire cubelet by the primary beam correction factor for the predicted central frequency and pixel of the source, and multiply the data by the distance (in Mpc) squared to convert the cube into line luminosity units.
    
    \item Produce an inverse variance weighted average stacked cube. In order to calculate the inverse variance weight, we use the cube before it is converted into line luminosity units. This is to avoid down-weighting the sources at the higher redshifts in the probed range. For a description of the different possible distance weights and their effects see \citet{delhaize13}.
    
    \item Produce an inverse variance weighted average point-spread function for the channels used in the stack. This is done extracting PSF cubelets in the same frequency ranges as the data cubelets and averaging them using the same weights as in the previous step.
    
    \item Perform a Hogbom CLEAN \citep{hogbom74} on the stacked cube using the average stacked point-spread function on the channels with expected emission. This is done by inspecting the cube and creating a mask over the identified emission. We then clean the flux in this mask down to the 1$\sigma$ level.
    
    \item Fit a second order polynomial to the channels identified as emission free and subtract it off. This is to remove any remaining imaging artifacts that may have been prevalent in the stack.
    
\end{enumerate}

\quad We chose to do a cubelet stacking, as opposed to spectral stacking, due to the fact that CHILES does resolve sources throughout our redshift range, and in order to produce an accurate HI measurement it is essential to CLEAN the stacked detection \citep{chen21a}. The emission of individual galaxies in the stack can not be cleaned due to the low signal to noise. As the stack raises the signal to noise it becomes essential to CLEAN the image. The HI mass for the stacked detection is calculated by summing the flux in the region with emission. We define the HI mass error as the standard deviation of a distribution of total flux fluxes in 100 emission-free regions from the HI stack cubelet.

\begin{figure}[t!]
\begin{center}
\includegraphics[width=\columnwidth]{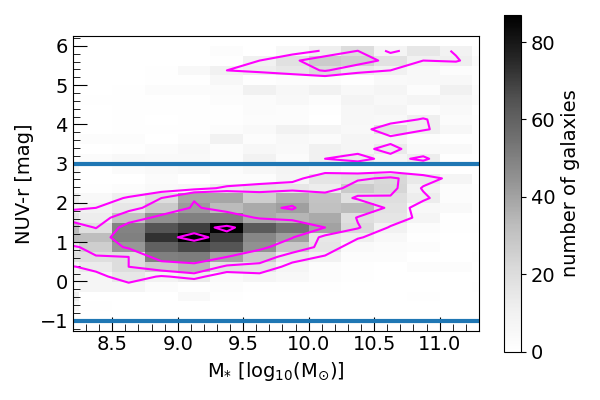}
\caption{The rest-frame NUV-r color as a function of stellar mass for all galaxies within the CHILES field. Both the color scale and the contours indicate the number of galaxies with each respective rest-frame NUV-r color and stellar mass combination. The solid blue lines indicate the boundaries of our blue color cutoff.}
\label{fig:ColorSelection}
\end{center}
\end{figure}

\begin{figure*}[t!]
\begin{center}
\includegraphics[width=\textwidth]{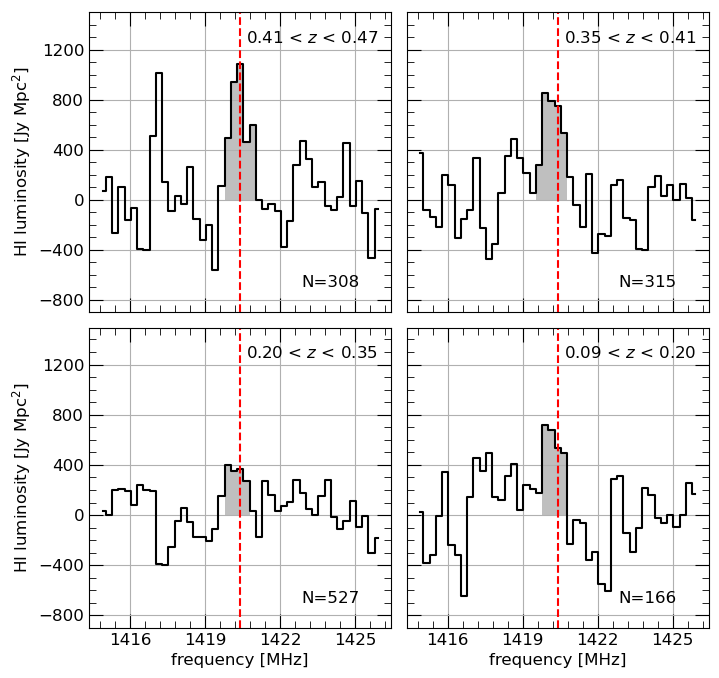}
\caption{The stacked HI spectra for blue galaxies in four redshift ranges continuous from 0.09 $ < z <$ 0.47. In each panel we indicate the redshift range, top right, number of galaxies included in the spectrum, bottom right, the rest frequency of HI (red dashed line), and grey shade the region integrated over for the HI mass calculation.}
\label{fig:HIstack_MassBlue}
\end{center}
\end{figure*}

\begin{figure*}[t!]
\begin{center}
\includegraphics[width=\textwidth]{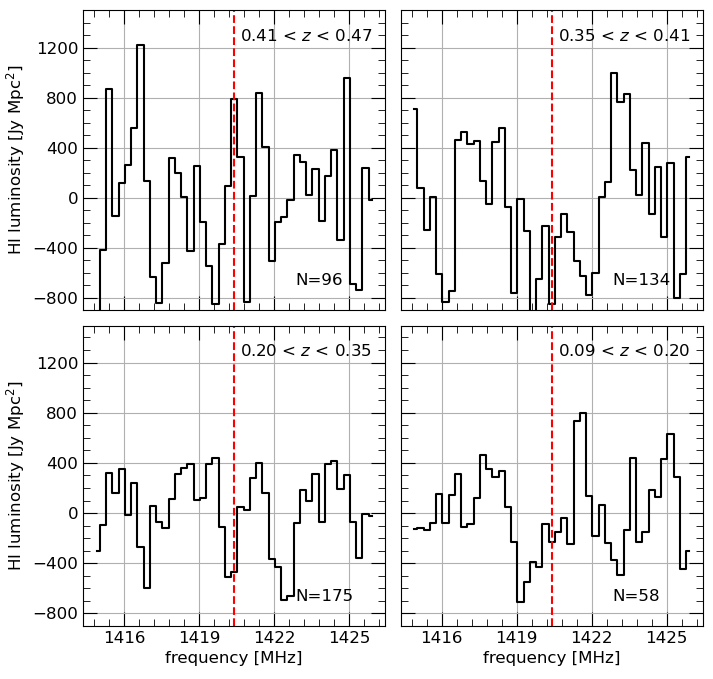}
\caption{The stacked HI spectra for red galaxies in four redshift ranges continuous from 0.09 $ < z <$ 0.47. In each panel we indicate the redshift range, top right, number of galaxies included in the spectrum, bottom right, and the rest frequency of HI (red dashed line),.}
\label{fig:HIstack_MassRed}
\end{center}
\end{figure*}

\quad In principle, our stacked measurement of average HI mass could be affected by unaccounted for galaxies that may lie within the beam of an input galaxy used in the stack. This effect is referred to as source confusion. In \citet{jones16}, the authors modeled the effect of stacked source confusion for several HI surveys, including CHILES, and they found that confusion introduces an error at the level of approximately 2 $\times$ 10$^{8}$ M$_{\odot}$. This is approximately an order of magnitude less that the HI masses we report and a factor 2-9 less than the error associated with the stacks, thus making it an insignificant source of error on our reported HI measurements. In addition, our high-resolution observations have the potential to spatially smear the stacked detection. This, and the fact that the stacked galaxies are initially unCLEANed, could have the effect of systematically lowering the measured HI mass.

\subsection{Radio Continuum Emission}\label{sec:cont_methodology}

\quad Radio continuum emission at 1.4 GHz is dominated by synchrotron emission which mostly arises from the effects of recent of recent star-formation or AGN. As is the case with radio spectral line data, one can stack radio continuum data for a sample of galaxies that are most likely to have the star-formation as the cause for radio continuum emission and derive an average flux density measurement and star-formation rate \citep{white07}. In order to calculate average star formation rates for the galaxies used in the HI stack, appropriately not including the galaxies that were rejected from the HI stack, we stack the radio continuum data from the CHILES Continuum Polarization Survey and use the scaling relationship between 1.4 GHz radio flux and star-formation rate in \citet{murphy11}. This technique allows us to calculate a statistical star-formation rate using the same techniques as our HI mass measurements ensuring that our reported values have been produced in a self-consistent manner.

\quad The manner in which we stack radio continuum is parallel to that of the HI stacking with a couple of key differences. First, as radio continuum is a single frequency intensity measurement, our stack comprises of two dimensional images, as opposed to the three dimensional cubelets from the HI stack. Second, in order to measure the parameter of interest, the rest 1.4 GHz emission, we scale the radio continuum flux of each source by using the following equation:
\begin{equation}
    F_{1.4GHz, rest} = F_{1.4GHz, obs} \times (1+z)^{\alpha}
\end{equation}\label{eq:flux}
where F$_{1.4GHz, obs}$ is the observed flux in the map, and alpha is taken to be -0.75, a value typical for radio continuum due to star-formation \citep{yun02}. Lastly, contrary to the HI data, the radio continuum image was CLEANED very deeply, a 5$\mu$Jy beam$^{-1}$ threshold (which is 3.5 times the rms noise level of 1.4 $\mu$Jy beam$^{-1}$). Our stack is by necessity a mix of cleaned and uncleaned sources, but since almost all continuum sources are unresolved at this resolution, we think this is an acceptable strategy. Aside from those two aspects, each step in the stack is identical, as well as the weighting schemes. 

\quad Using our stacked image, we can then measure the central flux and scale this value to calculate star-formation rates. This is done by leveraging the FIR-Radio correlation and the subsequent relationship between FIR flux and star-formation rates \citep{yun01,bell03}. Specifically, we use the following equation, Equation 17 in \citet{murphy11}:
\begin{equation}
    \left( \frac{S.F.R.}{M_{\odot}yr^{-1}} \right) = 6.35 \times 10^{-29} \left( \frac{L_{1.4GHz}}{erg s^{-1} Hz^{-1}} \right)
\end{equation}\label{eq:sfr}
for our calculation of average star-formation rate. One limitation of our procedure is that we do not reject radio flux that could be arising from AGN. We find that approximately 40\% of galaxies show at least some level of AGN activity, meaning that the radio flux is not necessarily dominated by AGN but rather AGN activity is present, which could systematically elevate our measured star-formation rates (Gim et al. in prep.). Additionally, our flux measurements could be limited by source confusion or blending, which would systematically raise our measurements, but are unlikely to be a major source of error due to the high sensitivity and resolution of CHILES Con Pol.

\begin{figure}
\begin{center}
\includegraphics[width=\columnwidth]{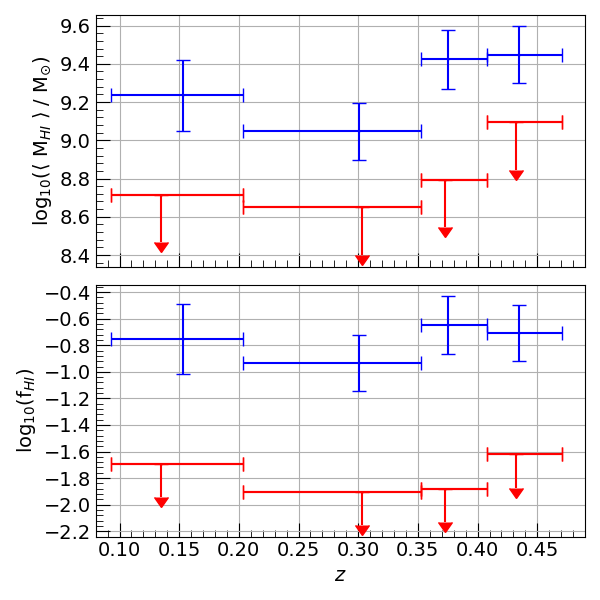}
\caption{The stacked HI mass (top) and HI gas fraction (bottom) as a function of redshift, for the blue galaxies (blue markers) and red galaxies (red markers). In all plots, the x-axis error bars correspond to the redshift bin boundaries, and the y-axis error bars indicate the errors calculated from the HI measurements. Galaxies that do not have an HI measurement are plotted as upper limits and are shown with downward pointing arrows.}
\label{fig:HIresults_ev}
\end{center}
\end{figure}

\begin{figure*}[t!]
\begin{center}
\includegraphics[width=\textwidth]{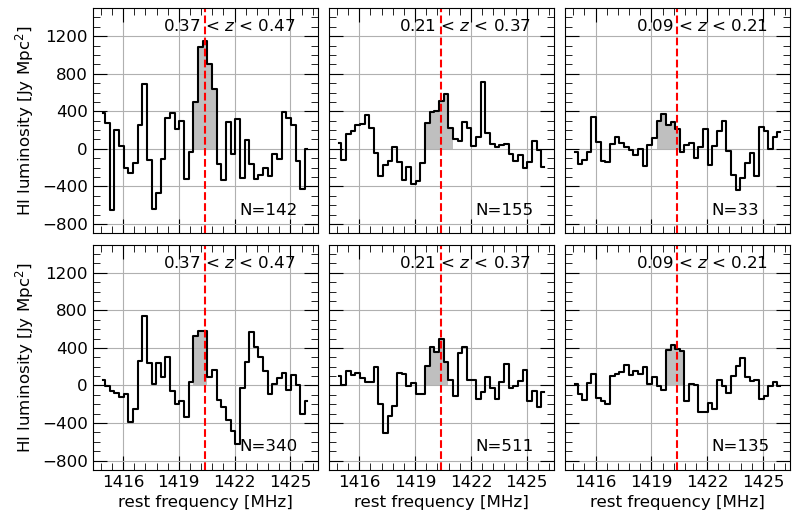}
\caption{The stacked HI spectra for blue galaxies in three redshift ranges continuous from 0.09 $ < z <$ 0.47, in the stellar mass range 10$^{10-12.5}$ M$_{\odot}$ (top) and 10$^{9-10}$ M$_{\odot}$ (bottom). In each panel we indicate the redshift range, top right, number of galaxies included in the spectrum, bottom right, the rest frequency of HI (red dashed line), and grey shade the region integrated over for the HI mass calculation.}
\label{fig:HIstack_MassDif}
\end{center}
\end{figure*}

\begin{figure*}[t!]
\begin{center}
\includegraphics[width=\textwidth]{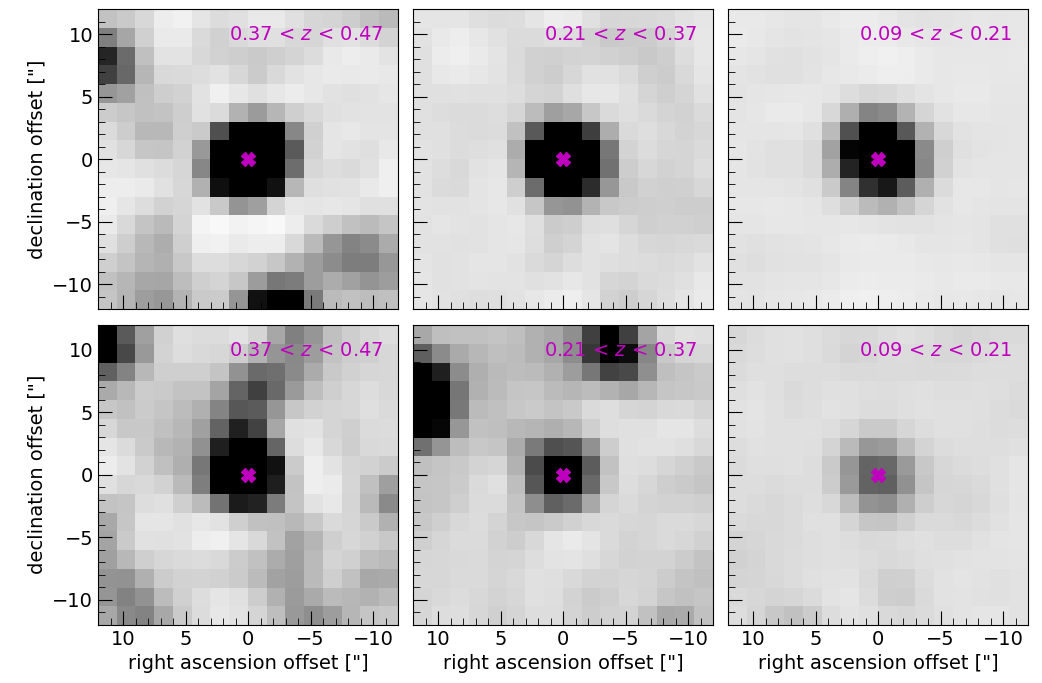}
\caption{The stacked radio continuum for blue galaxies in three redshift ranges continuous from 0.09 $ < z <$ 0.47, in the stellar mass range 10$^{10-12.5}$ M$_{\odot}$ (top) and 10$^{9-10}$ M$_{\odot}$ (bottom). In each panel we indicate the redshift range in the top right, and the color scale runs from -4 to 20 Jy Mpc$^{2}$ beam$^{-1}$ (top) and -3 to 8 Jy Mpc$^{2}$ beam$^{-1}$ (bottom). The magenta cross in the center indicates the average optical position for each galaxy in the stack.}
\label{fig:Contstack_MassDif}
\end{center}
\end{figure*}

\begin{figure}
\begin{center}
\includegraphics[width=\columnwidth]{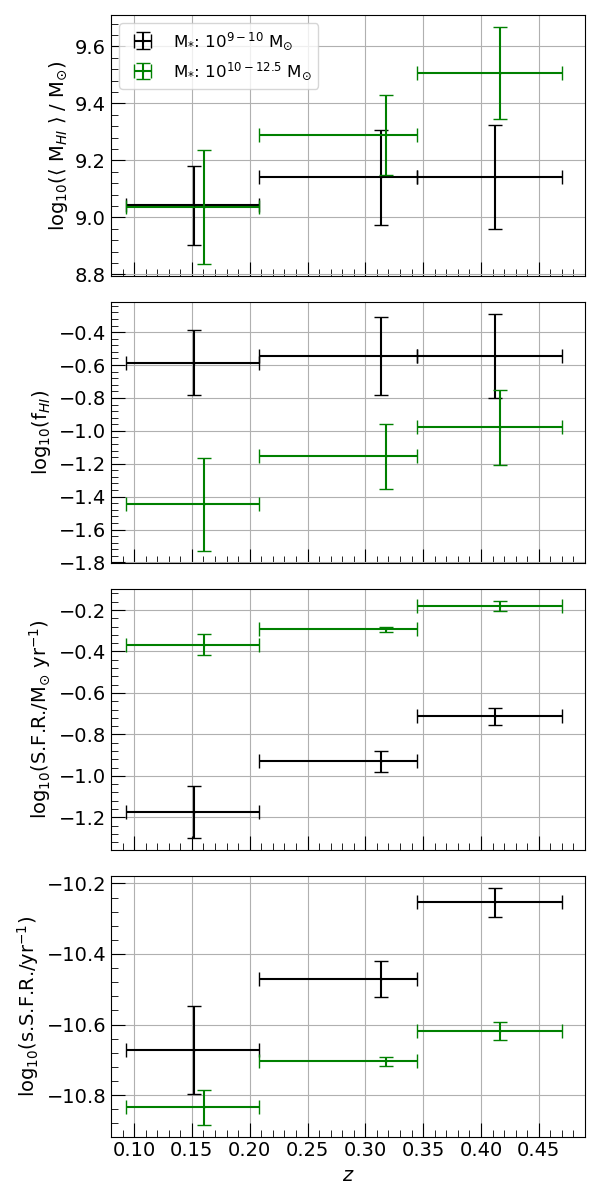}
\caption{The derived properties for blue galaxies in three redshift ranges continuous from 0.09 $ < z <$ 0.47, in the stellar mass ranges 10$^{10-12.5}$ M$_{\odot}$ (green) and 10$^{9-10}$ M$_{\odot}$ (black).}
\label{fig:HIres_MassDif}
\end{center}
\end{figure}

\begin{deluxetable*}{cccccccc}[t!]
\tablecaption{Stacking Results from CHILES\label{tab:res}}
\tablecolumns{7}
\tablenum{2}
\tablewidth{0pt}
\tablehead
{
\colhead{Redshift} &
\colhead{NUV-r} &
\colhead{N$_{gal}$} &
\colhead{Average HI Mass} &
\colhead{Average S.F.R.} &
\colhead{Average Stellar Mass} &
\colhead{Significance\tablenotemark{a}} &
\colhead{Relevant Figures} \\
\colhead{} & 
\colhead{mag} &
\colhead{} &
\colhead{10$^{9}$ M$_{\odot}$} &
\colhead{M$_{\odot}$ yr$^{-1}$} &
\colhead{10$^{9}$ M$_{\odot}$} &
\colhead{} &
\colhead{} 
}
\startdata
0.43 & -1 - 3  & 308 & 2.8$\pm$0.9 & 0.37$\pm$0.02 & 11.4  & 3.6  & 5 \& 7 \\
0.37 & -1 - 3  & 315 & 2.7$\pm$0.9 & 0.24$\pm$0.02 &  9.1  & 3.7  & 5 \& 7 \\
0.30 & -1 - 3  & 527 & 1.1$\pm$0.4 & 0.21$\pm$0.01 &  8.5  & 2.2  & 5 \& 7 \\
0.15 & -1 - 3  & 166 & 1.7$\pm$1.8 & 0.14$\pm$0.02 &  8.1  & 2.5  & 5 \& 7 \\ 
0.43 & 3 - 6   &  96 & $<1.3$      &            -  & 50.6  & -    & 6 \& 7 \\
0.37 & 3 - 6   & 134 & $<0.6$      &            -  & 46.8  & -    & 6 \& 7 \\
0.30 & 3 - 6   & 175 & $<0.4$      &            -  & 35.3  & -    & 6 \& 7 \\
0.13 & 3 - 6   &  58 & $<0.5$      &            -  & 25.1  & -    & 6 \& 7 \\
\hline
0.42 & -1 - 3  & 142 & 3.2$\pm$1.2 & 0.66$\pm$0.04 & 27.4  & 3.6  & 8 \& 9 \\
0.32 & -1 - 3  & 155 & 1.9$\pm$0.6 & 0.51$\pm$0.01 & 25.8  & 3.4  & 8 \& 9 \\
0.16 & -1 - 3  &  33 & 1.1$\pm$0.5 & 0.42$\pm$0.05 & 29.4  & 2.1  & 8 \& 9 \\ 
0.41 & -1 - 3  & 340 & 1.4$\pm$0.6 & 0.19$\pm$0.02 & 3.5   & 5.5  & 8 \& 9 \\
0.31 & -1 - 3  & 511 & 1.4$\pm$0.5 & 0.12$\pm$0.01 & 3.5   & 3.9  & 8 \& 9 \\
0.15 & -1 - 3  & 135 & 1.1$\pm$0.4 & 0.07$\pm$0.02 & 3.2   & 3.8  & 8 \& 9 \\ 
\enddata
\caption{The redshift ranges, color ranges, average galaxy properties, significances and relevant figures for the stacked detections presented in this work. Above the horizontal line is for the stacks that consider all galaxies with stellar masses greater than 10$^{9}$ M$_{\odot}$ and below the line are the results when we divide the blue galaxy sample into two stellar mass bins. When calculating average stellar masses and average star-formation rates, we only include the galaxies used in the HI stack and use the same weighting scheme as in the HI stack.}
\tablenotetext{a}{We define the significance of the HI stacked detection by dividing the peak flux in the spectra by the r.m.s. noise of the spectra, as measured in the line-free channels. These spectra correspond to the integrated HI profiles shown in Figures \ref{fig:HIstack_MassBlue} and \ref{fig:HIstack_MassDif}.}
\end{deluxetable*}

\section{Results} \label{sec:results}

\quad The wide bandwidth and sub 100 $\mu$Jy beam$^{-1}$ noise of the CHILES database processed in the manner described in Section \ref{sec:chiles_data}, combined with the HI stacking procedure outlined in Section \ref{sec:HI_methodology}, allow us to directly probe the average HI content of galaxies out to a redshift of 0.47, in contiguous redshift bins. All results presented in the following section are summarized in Table 2.

\subsection{Blue vs. Red Galaxies} \label{sec:results-color}

\quad To begin our investigation, we probe the evolution of average HI content for galaxies in four redshift bins, chosen to be approximately equal in redshift space and number of galaxies and without any a priori selection based on large-scale structure, spanning 0.09 $< z <$ 0.47. Specifically, in each bin we compare blue galaxies, defined as having a rest-frame NUV-r color in the range -1 to 3, and red galaxies, defined as having a rest-frame NUV-r color in the range 3 to 6, and stellar masses in the range 10$^{9-12.5}$ M$_{\odot}$. In Figure \ref{fig:ColorSelection}, we demonstrate how this color criterion separates the blue sequence versus the red sequence galaxies, allowing us to properly separate and investigate the evolution of the two different populations. Additionally, we choose our stellar mass criteria because COSMOS is complete for these stellar masses for our redshift bins \citep{davidzon17}. 

\quad In Figure \ref{fig:HIstack_MassBlue}, we show the stacked spectra for the blue galaxies, and in Figure \ref{fig:HIstack_MassRed} we show the stacked spectra for the red galaxies. For each of the blue galaxies, we detect and are able to determine an HI mass. However, for the red galaxies, each of our stacks resulted in a non-detection, and instead, an estimate for the HI mass upper limit. To produce each HI spectrum, we integrate over the equivalent of a 60 kpc box at the average redshift in each bin. In each panel the grey shaded region indicates the channels identified to have emission. This identification is determined by finding the first channels at both positive and negative velocities, with respect to restframe line center, that cross the expected one sigma value, and including all channels that lie between. These channels are then used to calculate the average HI mass. 

\quad With our stacked measurements in hand, we can now investigate the differences in HI content for galaxies over the past five billion years in red and blue galaxies. In Figure \ref{fig:HIresults_ev}, we show the HI mass (top) and HI gas fraction measurements (for the blue galaxies) and upper limits (for the red galaxies) as a function of redshift. The most obvious result from these measurements, is that the blue galaxies have a higher HI gas content, and that this relationship persists across the entire redshift range that we probe. Second, the HI content of blue galaxies with our stellar mass criterion do not significantly evolve in this redshift range. Each HI mass and HI gas fraction is reasonably within the error of the other measurements which indicates no significant change in the average HI gas content of massive blue galaxies in this redshift range. However, as a note, the HI measurement for blue galaxies in the second lowest redshift bin appears to scale with the noise. This is not surprising, given that this is amongst our lowest significance detection, and suggests that their may be some systematics affecting the data here. Acknowledgment of this caveat is further restricts us from over interpreting these results and claiming any evolution.

\subsection{The Effect of Stellar Mass} \label{sec:results-mass}

\quad The findings of the aforementioned section prompted a probe into the effect that mass may have on the redshift evolution of the HI content of blue galaxies. To explore this relationship, we stack blue galaxies, defined as having a NUV-r color in the range -1 to 3, in three redshift bins spanning 0.09 $< z <$ 0.47, in two different mass bins. For our two different mass bins, we selected two mass bins with cutoffs of stellar masses in the ranges 10$^{9-10}$ M$_{\odot}$ and 10$^{10-12.5}$ M$_{\odot}$ as our intermediate and high-mass mass bins, respectively. In Figure \ref{fig:HIstack_MassDif}, we present the stacked HI spectra, calculated in the same way as outlined above, for the blue galaxies in the high-mass bin (top) and intermediate mass bin (bottom). Additionally, in Figure \ref{fig:Contstack_MassDif}, we present the stacked radio continuum maps for the blue galaxies in the high-mass bin (top) and intermediate mass bin (bottom). The HI stacks, specifically the lowest redshift high mass bin and highest redshift low mass bin, are redshifted with respect to predicted linecenter. Both of these detections suffer from either negative features on the blue shifted side, low signal-to-noise, and comparable positive peaks. This does raise the possibility that these detections may be affected by systematic errors altering derived measurements. Additionally, low signal-to-noise ratio measurements suffer a larger uncertainty in the true measured linecenter further obfuscating the interpretation measurements of a velocity offset or width. As a result, below we concentrate specifically on trends across all redshift bins in order to gain an understanding on what these measurements mean.

\quad To investigate the possible differences in redshift evolution for our two stellar mass bins, in Figure \ref{fig:HIres_MassDif}, we present the HI mass, gas fraction, star-formation rate, and specific star-formation rate for the high mass bin (green markers) and intermediate mass bin (black markers) as a function of redshift. In these plots, we define $f_{HI}$ as the ratio of HI mass to stellar mass, and specific star-formation rate as the ratio of star-formation rate to stellar mass. These figures make clear that there is a clear difference in the redshift evolution of galaxies in these different stellar mass ranges. 

\quad To begin, let's focus on the redshift evolution of the average HI gas content. In the top panel of Figure \ref{fig:HIres_MassDif}, we see that for the intermediate mass bin there is little to no evolution of the HI gas mass, specifically, only an increase of a factor 1.3 that is not of statistical significance. For the high mass galaxies, we see a more noticeable increase of a factor 3.0, with the highest redshift bin measurement beyond the statistical errors of the lowest redshift bin, indicating a significant difference. For the HI gas fraction, there is a similar significant increase at the level of a factor 3.0 for the massive galaxies, and no significant change in the intermediate mass galaxy bin. This results show a clear decrease in HI mass for the most massive galaxies, and no trend for intermediate galaxies, in the redshift range, $z$ = 0.09 - 0.47.

\quad As for the star-formation rate evolution, we see the inverse relationship. The higher mass galaxies have consistently higher star formation rates, as is generally found for galaxies on the star-formation main sequence \citep{renzini15}. However, the relative redshift evolution of these galaxies shows some interesting variation. For the intermediate galaxies we see an increase of star-formation rate and specific star-formation rate by a factor of 2.9 and 2.6, respectively. For the high mass galaxies, we see an increase of star-formation rate and specific star-formation rate by a factor of just 1.5 and 1.6, respectively. 

\section{Discussion} \label{sec:disc}

\quad Our results demonstrate that the redshift evolution of the HI mass content of our sample of galaxies is not scaling with the the redshift evolution of the star-formation rate for these galaxies. This difference suggests that different physical mechanisms could be responsible for the regulation of star-formation rate for the different stellar masses we probe. The two clearest candidates for these mechanisms are some inefficiency in the HI to H$_{2}$ conversion, or some inefficiency in the ability of galaxies to replenish their HI gas reservoirs via accretion. In order to understand the specific role that HI has in the evolution of galaxies we will place our results within the context of the star-formation main sequence of galaxies.

\begin{figure}[t!]
\begin{center}
\includegraphics[width=\columnwidth]{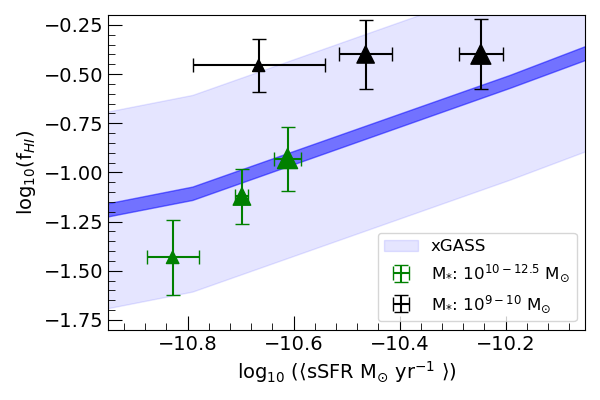}
\caption{The HI gas fraction as a function of specific star-formation rate for the galaxies in the CHILES field (individual points with error bars) and the galaxies in the xGASS survey (blue shaded regions). For the CHILES galaxies, the colors correspond to the two stellar mass ranges 10$^{10-12.5}$ M$_{\odot}$ (green) and 10$^{9-10}$ M$_{\odot}$ (black), and the sizes of the markers correlate with redshift, where larger marker size means higher redshift. The dark blue shaded region corresponds to the fit, and error on the fit, for the HI gas fraction as a function of specific star-formation rate, and the light blue shaded region indicates the typical scatter around this relationship, as reported in \citet{catinella18}.}
\label{fig:HIdisc_MassDif}
\end{center}
\end{figure}

\quad The star-forming main sequence of galaxies is the relationship between the star-formation rate and stellar masses of star-forming galaxies, and is often used as a benchmark by which one can study the evolution of galaxies \citep{speagle14,scoville17,ramatsoku20,bera23b}. For our study, we will look in particular at the relationship between HI gas fraction and specific star-formation rate. In order to compare our results to well known calibrations from the local universe, we will compare our measurements to those in \citet{catinella18} where they present fits to this relationship using the rich database of HI masses, stellar masses, and star-formation rates, as well as many other derived quantities, from the xGASS survey. In Figure \ref{fig:HIdisc_MassDif}, we show the HI gas fraction as a function of specific star-formation rate for the galaxies discussed in Section \ref{sec:results-mass}, where the colors correspond to the two different stellar mass bins and the marker size increases with increasing redshift. The results from \citet{catinella18} are shown as the blue shaded region, where the dark shaded region corresponds to the fitted relationship with errors as reported in \citet{catinella18} and the light shaded region corresponds to 0.5 dex from the mean, which is the characteristic scatter around the average value.

\quad Interestingly, the results in Figure \ref{fig:HIdisc_MassDif} are suggestive of different redshift evolution for these two samples of galaxies, albeit an evolution still within the scatter of the star-forming main sequence. The intermediate mass galaxies in the highest redshift bin are well within the expected value of the star-forming main sequence, and at decreasing redshift are becoming increasingly HI rich, as evident by the fact they are moving above the star-forming main sequence. For the high mass galaxies, we find that they similarly start well within the scatter of the star-forming main sequence, but at decreasing redshift are becoming increasingly HI poor, as evident by their moving below the star-forming main sequence. We point out that all of our data points lie within the scatter of the relationship presented in \citet{catinella18}, and that the interesting difference between the two mass bins lies in the differing gradients in redshift between the mass bins. This difference in gradient suggests both of these samples appear to be evolving in redshift, but in a manner opposite of one another.

\quad First, let's discuss the redshift evolution of the intermediate mass galaxies. Figures \ref{fig:HIres_MassDif} \& \ref{fig:HIdisc_MassDif} illustrate the blue galaxies in this mass range are maintaining their supply of cold atomic gas, but the star-formation rate is nonetheless declining. These two facts taken in conjunction illustrate that inefficient accretion from the circumgalactic and intergalactic media are not responsible for the decline in star-formation rate. Rather, it must be that the neutral hydrogen is not being effectively converted into molecular gas and subsequently formed into stars. For the high mass galaxies, Figures \ref{fig:HIres_MassDif} \& \ref{fig:HIdisc_MassDif} demonstrate that the blue galaxies in this mass range are failing to replenish their cold gas reservoirs. This suggests that these massive galaxies are not efficiently accreting gas from the circumgalactic and intergalactic media and subsequently not providing the raw fuel for conversion into molecular hydrogen and subsequently stars. However, for these high mass galaxies it could also be the case that the HI to H$_{2}$ conversion becoming less efficient, as is our proposed scenario for the intermediate mass galaxies, but our observations do not allow us to properly discriminate between these two effects for our results. 

\quad Several recent studies have used stacking techniques to study the evolution of HI mass content evolution out to these intermediate redshifts \citep[e.g.]{bera19,bera23a,bianchetti25arxiv}, and \citet{chowdhury22b} where they study HI content at redshifts out to 1.3. Additionally, in \citet{bera23b} and \citet{chowdhury23} study the gas accretion rate and molecule formation for redshifts out to 0.35 and 1.3, respectively. In both of these studies they use the respective HI mass to stellar mass scaling relationship they derive at different redshifts, known scaling relationships involving molecular gas content, and the assumption that galaxies evolve along the main sequence. Interestingly, \citet{chowdhury23} find that for their sample of galaxies it is a combination of lower net gas accretion and inefficient atomic to molecular conversion that drives the decrease in star-formation rate between redshift 1 and 0. In contrast, \citet{bera23b} find that in their redshift range, 0.35 - 0, is solely the inefficient conversion of atomic to molecular gas. In \citet{bianchetti25arxiv}, they derive the $M_{HI}-M_{*}$ relationship for exclusively star-forming galaxies at intermediate redshifts, and find that for this sample of galaxies it seems to be the case that HI is not being accreted quickly enough to replenish their gaseous reservoirs. Within this context, our results do not prove or disprove either scenario, but instead show there is a difference in the evolution of HI content of galaxies of different masses, and different physical scenarios can explain this evolution. Therefore, in order to confidently discuss the evolution of galaxies, we need to separate them by stellar mass and star-formation rate in order to form a coherent picture.

\quad As we begin to have meaningful statistical results of average HI content out to higher redshifts there are several different manners in which more observational data can help us understand the answer to our posed question. First, and most obviously, is the direct detection of neutral atomic and molecular gas content in individual galaxies across redshift space. Although current surveys, such as CHILES, Looking At the Distant Universe with the Meerkat Array\citep[LADUMA,][]{blyth16}, MIGHTEE-HI \citep{maddox21}, will be able to contribute to the literature on this front, they will lack the numbers of galaxies necessary to provide a robust statistical basis to prove these claims. Such direct detections in neutral hydrogen will likely have to wait until the era of the Square Kilometer Array. Additionally, large-scale dust mass and H$_{2}$ surveys will complement these upcoming SKA surveys by contributing to the total knowledge of the baryon cycle for these galaxies. However, this study demonstrates that future surveys should have galaxies of all masses as the baryonic physics taking place within them that is responsible for the decline in star-formation may differ between masses.

\section{Summary} \label{sec:sum}

\quad In this work, we have utilized the full CHILES survey to investigate the evolution of the average HI content of galaxies over the continuous redshift range 0.09 $< z <$ 0.47. In order to do this, we introduced a fast imaging pipeline that utilizes a per-observing session multi-step and multi-scale \textit{uv}-plane based subtraction. We then demonstrate the applicability and success of \textit{uv}-plane based gridding and averaging, which we perform per observing epoch, to greatly reduce our total visibility data volume that is imaged. An image including all of the processed CHILES data then is passed through a gridded \textit{uv}-plane RFI mitigation technique which successfully removes large-scale artifacts from our data. The effect of this procedure is then quantified through discussion of kurtosis and noise values which allows us to conclude that the CHILES data has been successfully processed through this fast imaging pipeline and is suitable for science.

\quad Using the calibrated, imaged, and continuum subtracted CHILES data, as well as ancillary data from the COSMOS collaboration and CHILES Con Pol, we measure average HI mass, HI gas faction, and star-formation rate using stacking techniques. The results of these measurements are summarized as follows:

\begin{enumerate}

    \item For the total ensemble of massive blue galaxies, which we define as galaxies with stellar masses in the range 10$^{9 - 12.5} M_{\odot}$, and an NUV-r color in the range -1 to 3, we find little-to-no evolution in the HI gas content for galaxies in the redshift range 0.09 $< z <$ 0.47.

    \item For the massive red galaxies, which we define as galaxies with stellar masses in the range 10$^{9 - 12.5} M_{\odot}$, and an NUV-r color in the range 3 to 6, we make no HI mass detection at any redshift indicating that it is not the case that red galaxies are becoming significantly HI rich at increasing redshift such that they would be detected.

    \item We separated our sample blue galaxies into two mass bins, an intermediate mass bin, comprised of galaxies with stellar masses in the range 10$^{9 - 10} M_{\odot}$ and high mass bin, comprised of galaxies with stellar masses in the range 10$^{10 - 12.5} M_{\odot}$. For the intermediate stellar mass bin, we find no significant evolution in the HI gas content. However, the high stellar mass bin shows significant increase in HI content for increasing redshifts.

    \item We place our results into the context of the star-forming main sequence of galaxies and hypothesize that for the intermediate galaxies the decline in star-formation rate is caused by an inefficient conversion of atomic to molecular hydrogen. For the high mass galaxies, we find that HI is not being accreted onto the disks of these galaxies, which could be the cause of a decline in star-formation rate, as well as an inefficient conversion of atomic to molecular hydrogen
    
\end{enumerate}

\quad The results presented here add to the recent findings of multiple groups using multiple different telescopes to understand the evolution of HI at intermediate redshifts using stacking techniques \citep[e.g.][]{bera19,chowdhury20,bera23a,bianchetti25arxiv}. As new deep long-integration HI surveys are being undertaken \citep[e.g. LADUMA][]{blyth16}, future studies will be able to more properly fill in the redshift space of $z$ = 0.5 - 2, and add to the results from \citet{chowdhury20}, allowing for us to fully appreciate the role of HI in the evolution of galaxies. Additionally, next generation telescopes will provide the spatial resolution neccessary in order to understand the sundry physical processes that are governing the star-formation rate evolution of the Universe.

\begin{acknowledgments}

\quad We gratefully acknowledge the thorough review and helpful comments from the anonymous referee that added to the clarity, demonstration of results, and scientific discussion, all of which served to greatly improved this paper.

\quad We greatly appreciate valuable insights from Nick Scoville on an earlier draft of this work concerning the evolution of the total ISM. Additionally, we are grateful to Kumar Golap for helpful instructions on the usage of \textit{msuvbin}, to Bradley Frank for helpful conversations on the parallelization of \textit{tclean}. Lastly, we thankfully acknowledge Guillermo Avendano Franco and Nate Garver-Daniels for their invaluable assistance with the use of the WVU HPCs.

The National Radio Astronomy Observatory is a facility of the National Science Foundation operated under cooperative agreement by Associated Universities, Inc. The CHILES survey was partially supported by a collaborative research grant from the National Science Foundation under grant Nos. AST—1412843, 1412578, 1413102, 1413099, and 1412503.

\quad DJP greatfully acknowledges support by the South African Research Chairs Initiative of the Department of Science and Technology and the National Research Foundation. Julia Blue Bird was a Jansky Fellow of the National Radio Astronomy Observatory.

\end{acknowledgments}

\vspace{5mm}

\facilities{EVLA \citep{evla}, WVU High Performance Computing}

\software{astropy \citep{astropy}, CASA \citep{casa}}

\bibliography{references}
\bibliographystyle{aasjournal}

\newpage

\begin{appendix} 

\section{Stacked Measurements}

\quad In the following figure, we present plots indicating the reliability for each of our stacked detections. Specifically, we show the stacked spectrum overlaid on 100 blank sky spectra, and a comparison of how our the noise in our stacked spectra integrates down, compared to the expected N$^{-0.5}$, where N is the number of input galaxies for the stack. These figures demonstrate that, on average, features in the spectra are representative of global per-channel image artifacts, as opposed to purely serendipitous or spurious features, and that the noise for our stacks broadly agrees with expectations. The spectra here are ordered by their order of appearance in the paper.

\begin{figure}[h!]
\includegraphics[width=0.98\textwidth]{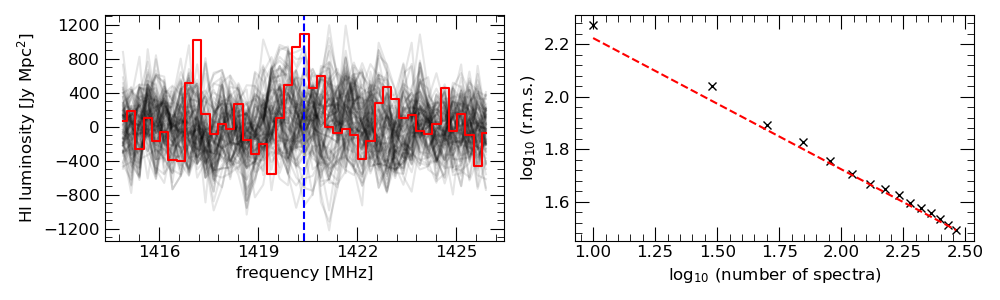}
\end{figure}
\begin{figure}[h!]
\includegraphics[width=0.98\textwidth]{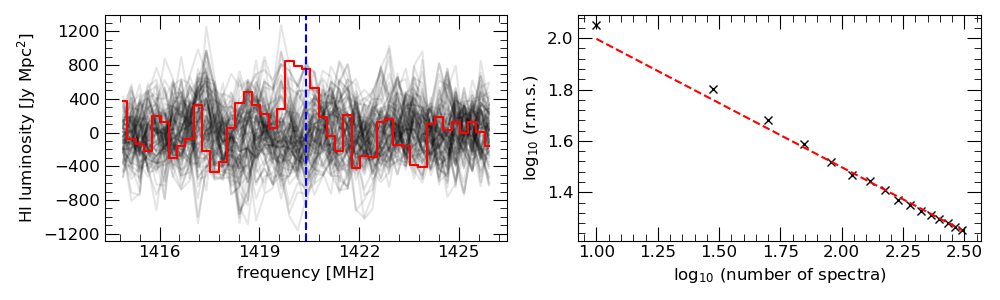}
\end{figure}
\begin{figure}[h!]
\includegraphics[width=0.98\textwidth]{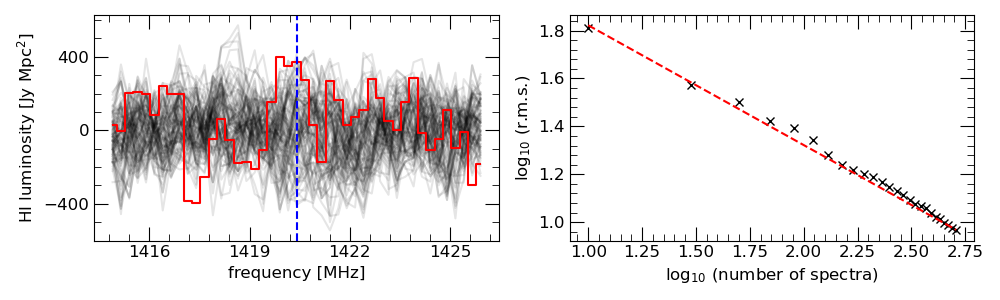}
\end{figure}
\begin{figure}
\includegraphics[width=0.98\textwidth]{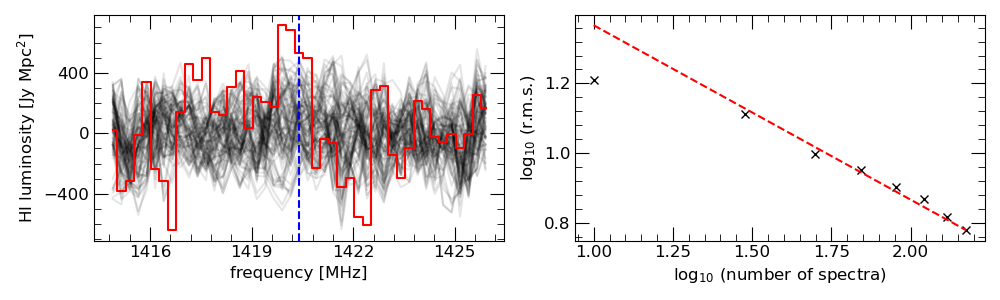}
\end{figure}
\begin{figure}
\includegraphics[width=0.98\textwidth]{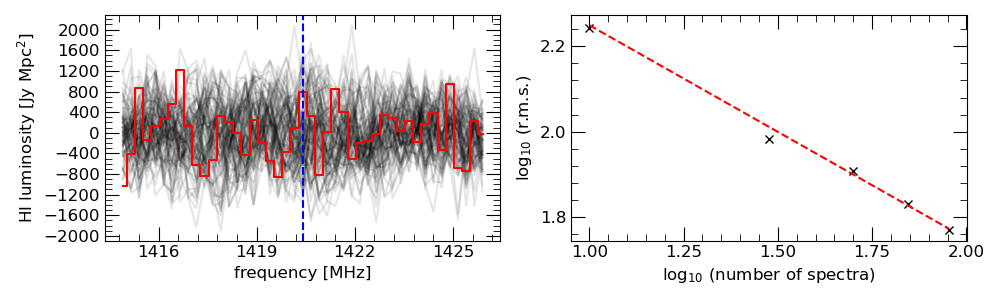}
\end{figure}
\begin{figure}
\includegraphics[width=0.98\textwidth]{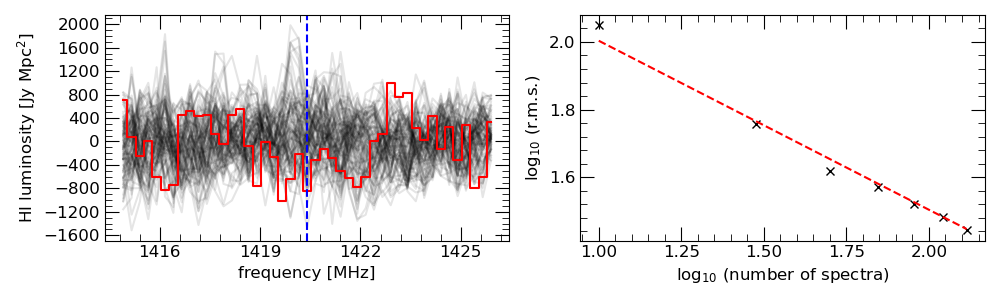}
\end{figure}
\begin{figure}
\includegraphics[width=0.98\textwidth]{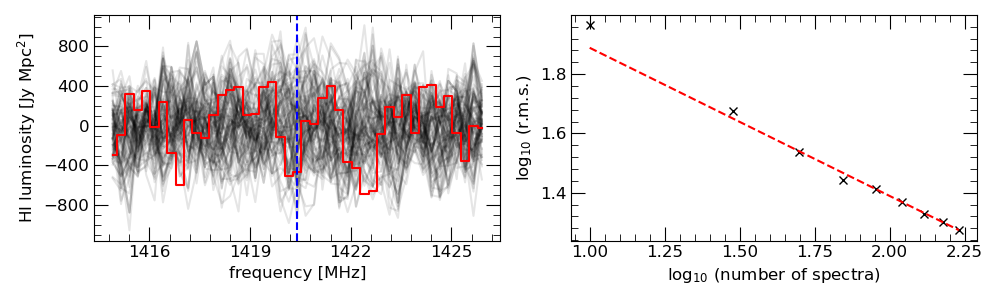}
\end{figure}
\begin{figure}
\includegraphics[width=0.98\textwidth]{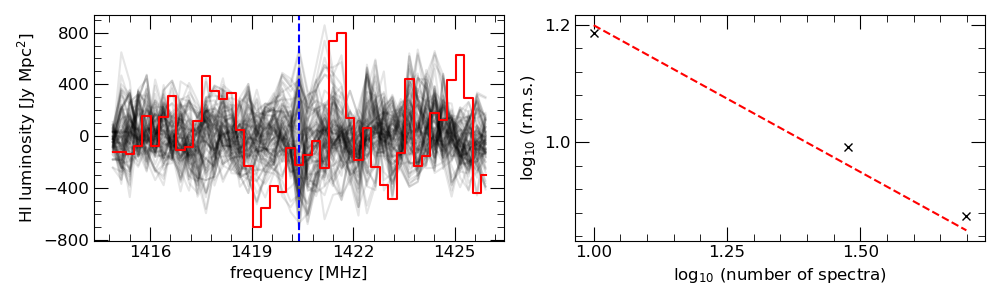}
\end{figure}
\begin{figure}
\includegraphics[width=0.98\textwidth]{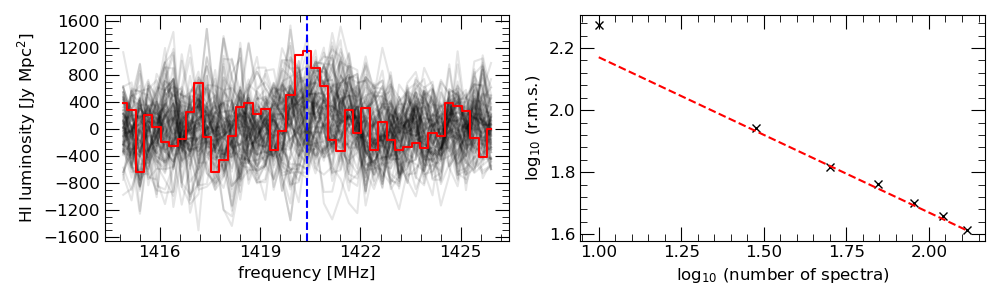}
\end{figure}
\begin{figure}
\includegraphics[width=0.98\textwidth]{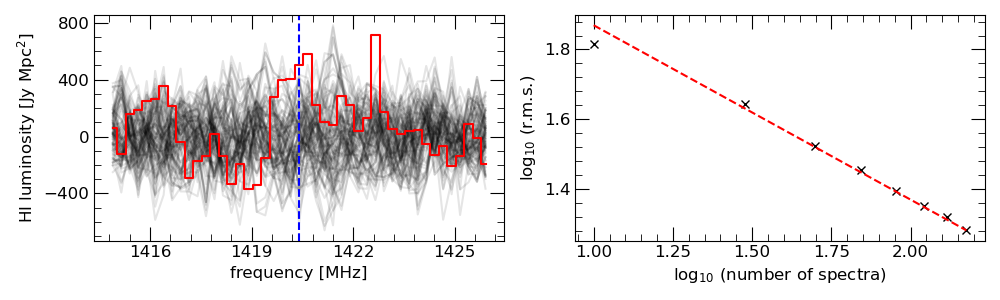}
\end{figure}
\begin{figure}
\includegraphics[width=0.98\textwidth]{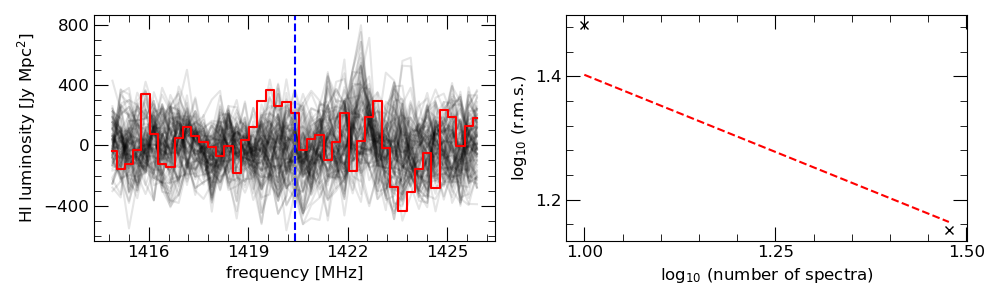}
\end{figure}
\begin{figure}
\includegraphics[width=0.98\textwidth]{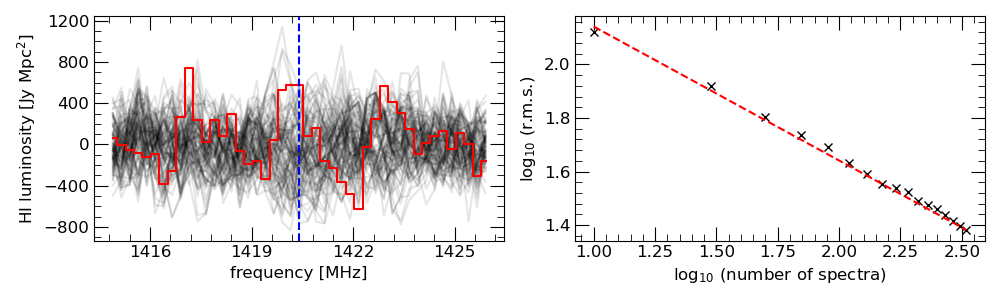}
\end{figure}
\begin{figure}
\includegraphics[width=0.98\textwidth]{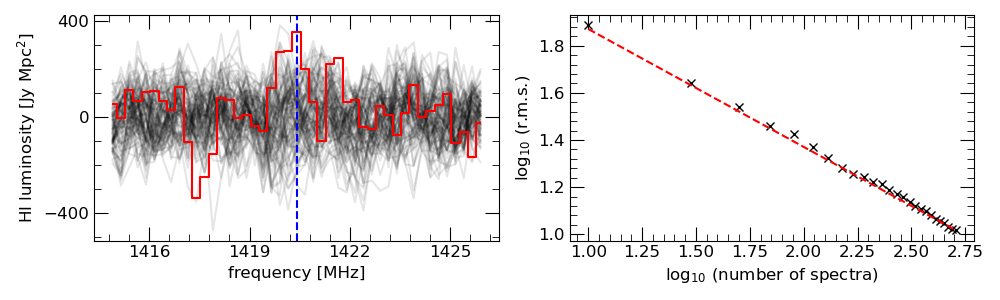}
\end{figure}
\begin{figure}
\figurenum{A.1}
\includegraphics[width=0.98\textwidth]{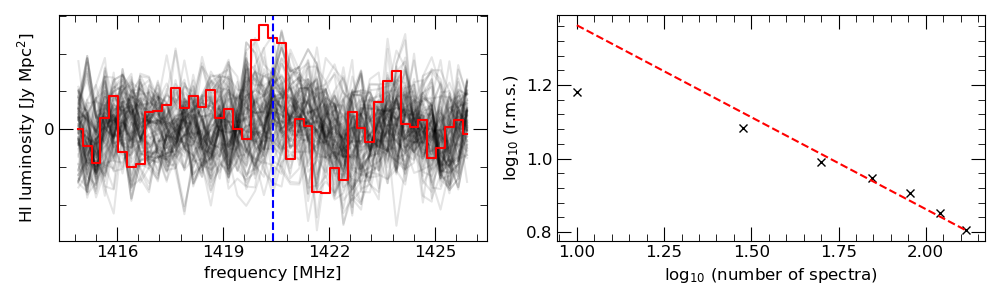}
\caption{For each of the 14 stacked spectra summarized in Table \ref{tab:res} we present: \textit{\textbf{left:}} the stacked HI spectrum (red), 100 reference stacked spectra taken over randomized locations in the stacked cube (black), and the HI line rest frequency (blue dashed), and \textit{\textbf{right:}} the observed rms noise (black markers) and expected rms noise, assuming an N$^{-0.5}$ noise integration and the final stacked noise as the anchoring point (red dashed).}
\label{fig:appendix}
\end{figure}

\end{appendix}

\end{document}